\documentclass[aps,prd,amsmath,twocolumn,showpacs,amssymb]{revtex4}
\newcommand{\be}{\begin{eqnarray}}
\newcommand{\ee}{\end{eqnarray}}
\newcommand{\nn}{~\nonumber\\}
\newcommand{\sint}{{\textstyle\int}}
\newcommand{\sfrac}[2]{{\textstyle\frac{#1}{#2}}}
\newcommand{\M}{\mathbbm{M}}
\newcommand{\Det}{\mathrm{Det}}
\newcommand{\tr}{\mathrm{tr}}
\usepackage{bbm}
\newcommand{\B}{\mathbbm{B}}
\newcommand{\m}{{\bf m}}
\newcommand{\afluc}{a}
\newcommand{\asad}{\breve A}
\newcommand{\adj}{\mathring}
\newcommand{\g}{\mathrm{g}}
\newcommand{\im}{{\cal I}}
\newcommand{\re}{{\cal R}}
\newcommand{\ul}[1]{\underline{#1}}

\begin{document}

\title{Electroweak symmetry breaking in other terms}

\author{Dennis D.~Dietrich}

\affiliation{Institut for Fysik og Kemi, Syddansk Universitet, Odense,
Danmark}

\date{April 7, 2008}

\begin{abstract}
We analyse descriptions of electroweak symmetry breaking in terms of 
ultralocal antisymmetric tensor fields and gauge-singlet geometric
variables, respectively; in particular, the Weinberg--Salam model and,
ultimately, dynamical electroweak symmetry breaking by technicolour theories
with enhanced symmetry groups. Our motivation is to unveil the manifestly
gauge invariant structure of the different realisations. We find, for
example, parallels to different types of torsion.
\end{abstract}


\maketitle

\section{Introduction}

Massive gauge bosons belong to the fundamental concepts we use for picturing
nature. Apart from electroweak symmetry breaking, which here is our main
interest, other prominent examples are superconductivity and confinement. The Drosophila for electroweak symmetry breaking is the Weinberg--Salam model which is at the basis of the standard model. It relies on breaking the electroweak symmetry by the coupling to an elementary scalar particle, the Higgs. Theoretical shortcomings of the basic model provide the incentive for looking for modifications beyond the standard model. One of those, which shall be studied here is known as technicolour \cite{TC}. In technicolour the electroweak symmetry is broken by chiral symmetry breaking among fermions (techniquarks) in an additional strongly interacting sector which replaces the Higgs sector in the standard model.

Even more than in the massless case, gauge invariance is a severe constraint
for the construction of massive gauge field theories. Usually, additional
fields beyond the original gauge field have to be included in order to
obtain gauge invariant expressions. Technically, this is linked to the fact
that the gauge field changes inhomogeneously under gauge transformations and
encodes also spurious degrees of freedom arising from the construction
principle of gauge invariance. All this complicates the extraction of
physical quantities and might shroud essentials of the physical content of the theory.
A variety of approaches has been developed in order to deal with this
situation. Wilson loops \cite{Wilson}, for example, represent gauge invariant but
non-local variables.
Alternatively, there exist 
decomposition techniques like the one due to Cho, Faddeev, and Niemi
\cite{CFN}. Here, we first pursue a reformulation \footnote{There are
other descriptions of massive theories involving antisymmetric tensor fields
not arising from a reformulation \cite{lahiri}.} 
in terms of antisymmetric gauge algebra valued tensor fields 
$B^a_{\mu\nu}$ (Sect.~\ref{FSR}) and subsequently continue with a
representation in terms of geometric variables (Sect. \ref{GEOM}).

Before we treat the Weinberg--Salam model and technicolour models in Sect.~(\ref{NONDIAG}), we begin our discussion with the less involved set-ups of massless and massive field theories with simple gauge groups.
In Sect.~\ref{FSR0} we review the massless case. The antisymmetric tensor field
transforms homogeneously under gauge transformations. 
This fact already makes it simpler to account for gauge invariance. In
Sect.~\ref{FSRM} the generalisation to the massive case is presented. In the
$B^a_{\mu\nu}$ field representation the (non-Abelian) St\"uckelberg fields,
which are commonly present in massive gauge field theories and needed there in
order to keep track of gauge invariance, factor out completely. In other
words, no scalar fields are necessary for a gauge invariant formulation of
massive gauge field theories in terms of antisymmetric tensor
fields. The case of a constant mass is linked to sigma models (gauged and ungauged) in different 
respects. Sect.~\ref{FSRH} contains the
generalisation to a position dependent mass, which corresponds to
introducing the Higgs degree of freedom. In Sect.~\ref{NONDIAG} non-diagonal
mass terms are admitted. This is necessary to accommodate the electroweak symmetry breaking pattern. The Weinberg--Salam model and technicolour models are studied as particular cases. Technicolour models with enhanced symmetry groups have additional ``pions'' beyond the ones corresponding to the aforementioned
St\"uckelberg degrees of freedom.

Sect.~\ref{GEOM} presents a description of the massive case in terms of geometric variables. In this step the remaining gauge degrees of freedom are eliminated. The resulting description is in terms of local colour singlet variables. Finally, Sect.~\ref{GEOMWS} is
concerned with the geometric representation of the Weinberg--Salam model and Sect.~\ref{GEOMTC} with that of technicolour. In Sect.~\ref{TOR}, we make the link between the different variants of symmetry breaking and different contributions to torsion.

The Appendix treats the Abelian case. It allows to better
interpret and understand several of the findings in the non-Abelian
settings. Of course, in the Abelian case already the $B_{\mu\nu}$ field is
gauge invariant. One also sees that the $m\rightarrow
0$ limit of the gauge propagator for the $B_{\mu\nu}$ fields is well-defined
as opposed to the ill-defined limit for the $A_\mu$ field propagator.

Sect.~\ref{SUM} summarises the paper.

\section{Antisymmetric tensor fields\label{FSR}}
\subsection{Massless\label{FSR0}}
Before we investigate massive gauge field theories let us recall some
details about the massless case. The partition function of a massless
non-Abelian gauge field theory without fermions is given by
\be
P:=\int[dA]\exp\{i\sint d^4x{\cal L}\},
\label{part1}
\ee
with the Lagrangian density
\be
{\cal L}={\cal L}_0:=-\sfrac{1}{4g^2}F^a_{\mu\nu}F^{a\mu\nu}
\ee
and the field tensor
\be
F^a_{\mu\nu}
:=
\partial_\mu A^a_\nu-\partial_\nu A^a_\mu+f^{abc}A^b_\mu A^c_\nu.
\ee
$A^a_\mu$ stands for the gauge field, $f^{abc}$ for the antisymmetric
structure constant, and $g$ for the coupling constant. \footnote{The
functional integral over a gauge field is ill-defined as long as no gauge is
fixed. We have to keep this fact in mind at all times and will discuss
it in detail when a field is really integrated out.} Variation of the
classical action with respect to the gauge field gives the classical
Yang--Mills equations
\be
D^{ab}_\mu(A) F^{b\mu\nu}=0,
\label{YM}
\ee
where the covariant derivative is defined as 
$D^{ab}_\mu(A):=\delta^{ab}\partial_\mu+f^{acb}A^c_\mu$.
The partition
function in the first-order formalism can be obtained after multiplying
Eq.~(\ref{part1}) with a prefactor in form of a Gaussian integral over an
antisymmetric tensor field $B^a_{\mu\nu}$,
\be
P
&\cong&
\int[dA][dB]
\exp\{i\sint d^4x[{\cal L}_0
-\sfrac{g^2}{4}B^a_{\mu\nu}B^{a\mu\nu}]\}.
\ee
("$\cong$" indicates that in the last step the normalisation of the partition
function has been changed.)
Subsequently, the field $B^a_{\mu\nu}$ is shifted by $\frac{1}{g^2}\tilde
F^a_{\mu\nu}$, where the dual field tensor is defined as 
$\tilde F^a_{\mu\nu}
:=
\frac{1}{2}\epsilon_{\mu\nu\kappa\lambda}F^{a\kappa\lambda}$, 
\be
P
&=&
\int[dA][dB]
\times
\nn
&&\times
\exp\{i\sint d^4x[
-\sfrac{1}{2}\tilde F^a_{\mu\nu}B^{a\mu\nu}
-\sfrac{g^2}{4}B^a_{\mu\nu}B^{a\mu\nu}]\}.
\label{part2}
\ee
In this form the partition function is formulated in terms of the
Yang--Mills connection $A^a_\mu$ and the antisymmetric tensor field
$B^a_{\mu\nu}$ as independent variables. Variation of the classical action with
respect to these variables leads to the classical equations of motion
\be
g^2B^a_{\mu\nu}
=
-
\tilde F^a_{\mu\nu}~~~\mathrm{and}~~~D^{ab}_\mu(A)\tilde B^{b\mu\nu}=0,
\label{cleoms}
\ee
where 
$\tilde B^a_{\mu\nu}
:=
\frac{1}{2}\epsilon_{\kappa\lambda\mu\nu}B^{a\kappa\lambda}$.
By eliminating $B^a_{\mu\nu}$ the original Yang--Mills equation (\ref{YM})
is reproduced. Every term in the classical action in the partition function 
(\ref{part2}) contains at most one derivative as opposed to two in
Eq.~(\ref{part1}). This explains the name "first-order" formalism. The
classical action in Eq.~(\ref{part2}) is invariant under simultaneous gauge
transformations of the independent variables according to
\be
A^{a\mu}T^a&=:&A^\mu\rightarrow A^\mu_U:=U[A^\mu-iU^\dagger(\partial^\mu U)]U^\dagger\\
B^{a\mu\nu}T^a&=:&B^{\mu\nu}\rightarrow B^{\mu\nu}_U:=UB^{\mu\nu}U^\dagger,
\label{bigtrafo}
\ee
or infinitesimally,
\be
\delta A^a_\mu&=&\partial_\mu\theta^a+f^{abc}A^b_\mu\theta^c\nn
\delta B^a_{\mu\nu}&=&f^{abc}B^b_{\mu\nu}\theta^c.
\label{homogtrafo}
\ee
The $T^a$ stand for the generators of the gauge group.
From the Bianchi identity $D_\mu^{ab}(A)\tilde F^{b\mu\nu}=0$ follows a second
symmetry of the $BF$ term alone: Infinitesimally, for unchanged $A^a_\mu$, 
\be
\delta B^a_{\mu\nu}
=
\partial_\mu\vartheta_\nu^a
-
\partial_\nu\vartheta_\mu^a
+
f^{abc}(A^b_\mu\vartheta^c_\nu-A^b_\nu\vartheta^c_\mu).
\label{dualtrafo}
\ee
A particular combination of the transformations (\ref{homogtrafo}) and 
(\ref{dualtrafo}),
$\theta^a=n^\mu A_\mu^a$
and
$\vartheta^a_\nu=n^\mu B^a_{\mu\nu}$,
corresponds to the transformation of a tensor under an infinitesimal local 
coordinate transformation $x^\mu\rightarrow x^\mu-n^\mu(x)$,
\be
\delta B_{\mu\nu}
=
B_{\lambda\nu}\partial_\mu n^\lambda
+
B_{\mu\lambda}\partial_\nu n^\lambda
+
n^\lambda\partial_\lambda B_{\mu\nu},
\ee 
that is a diffeomorphism.
Hence, the $BF$ term is diffeomorphism invariant, which explains why 
this theory is also known as $BF$ gravity. The $BB$ term is not diffeomorphism
invariant and, hence, imposes a constraint. The combination of the two terms
amounts to an action of Plebanski type which are studied in the context of
quantum gravity \cite{MacMan,BF}.

We now would like to eliminate the Yang--Mills connection by integrating it 
out. For fixed $B^a_{\mu\nu}$ the integrand of the path integral is not
gauge invariant with respect to gauge transformations of the gauge field
$A^a_\mu$ alone; the field tensor $F^a_{\mu\nu}$ transforms homogeneously and
the corresponding gauge transformations are not absorbed if $B^a_{\mu\nu}$
is held fixed. Therefore, the integral over the gauge group is in general not 
cyclic which otherwise would render the path integral ill-defined. The term in the 
exponent linear in the gauge field $A^a_\mu$, 
$A^a_\nu\partial_\mu\tilde B^{a\mu\nu}$,
is obtained by carrying out a partial integration in which surface terms are 
ignored. Afterwards it is absorbed by shifting $A^a_\mu$ by 
$(\B^{-1})^{ab}_{\mu\nu}(\partial_\lambda\tilde B^{b\lambda\nu})$,
where $\B^{ab}_{\mu\nu}:=f^{abc}\tilde B^a_{\mu\nu}.$ In general its inverse
$(\B^{-1})^{ab}_{\mu\nu}$, defined by 
$(\B^{-1})^{ab}_{\mu\nu}\B^{bc}_{\kappa\lambda}g^{\nu\kappa}
=
\delta^{ac}g_{\mu\lambda}$
exists in three or more space-time dimensions \cite{DT}
\footnote{Singular configurations can be linked to the Wu--Yang
ambiguity \cite{WY}.}. We are left with a
Gaussian integral in $A^a_\mu$ giving the inverse square-root of the 
determinant of $\B^{ab}_{\mu\nu}$,
\be
\Det^{-\frac{1}{2}}\B
&:=&
\prod_x{\det}^{-\frac{1}{2}}\B
\cong
\nn
&\cong&
\int[d\afluc]\exp\{-\sfrac{i}{2}\sint d^4x 
\afluc^{a\mu}\B^{ab}_{\mu\nu}\afluc^{b\nu}\}. 
\label{det}
\ee
In the last expression $\B^{ab}_{\mu\nu}$ appears in the place of an
inverse gluon propagator, that is sandwiched between two gauge fields. This
analogy carries even further: Interpreting $\partial_\mu\tilde B^{a\mu\nu}$
as a current, 
$(\B^{-1})^{ab}_{\mu\nu}(\partial_\lambda\tilde B^{b\lambda\nu})$, the
current together with the "propagator" $(\B^{-1})^{ab}_{\mu\nu}$, is exactly
the abovementioned term to be absorbed in the gauge field $A^a_\mu$.
Finally, we obtain,
\be
P
&\cong&
\int[dB]
\Det^{-\frac{1}{2}}\B
\exp\{i\sint d^4x[-\sfrac{g^2}{4}B^a_{\mu\nu}B^{a\mu\nu}
-
\nn
&&-
\sfrac{1}{2}
(\partial_\kappa\tilde B^{a\kappa\mu})
(\B^{-1})^{ab}_{\mu\nu}
(\partial_\lambda\tilde B^{b\lambda\nu})
]\}.
\label{massless}
\ee
This result is known from \cite{DT,H,GS}.
The exponent in the previous expression corresponds to the value of the $[dA]$
integral at the saddle-point value $\asad_\mu^a$ of the gauge field.
It obeys the classical field equation (\ref{cleoms}). Using 
$\asad_\mu^a(B)
=
(\B^{-1})^{ab}_{\mu\nu}(\partial_\lambda\tilde B^{b\lambda\nu})$ the
second term in the above exponent can be rewritten as
$-\sfrac{i}{2}\int d^4x\tilde B^a_{\mu\nu}F^{a\mu\nu}[\asad(B)]$,
which involves an integration by parts and makes its gauge invariance
manifest. The fluctuations $\afluc^a_\mu$ around the saddle point $\asad_\mu^a$,
contributing to the partition function (\ref{part2}), are Gaussian because the 
action in the first-order formalism is only of second order in the gauge 
field $A^a_\mu$. They give rise to the determinant (\ref{det}). 
What happens if a zero of the determinant is encountered can be understood by
looking at the Abelian case discussed in Appendix \ref{appa}. There the $BF$
term does not fix a gauge for the integration over the gauge field $A_\mu$
because the Abelian field tensor $F^{\mu\nu}$ is gauge invariant. 
If it is performed nevertheless one encounters a functional $\delta$
distribution which enforces the vanishing of the current 
$\partial_\mu\tilde B^{\mu\nu}$. In this sense the zeros of the determinant 
in the non-Abelian case arise if $\tilde B^a_{\mu\nu}$ is such that the $BF$ 
term does not totally fix a gauge for the $[dA]$ integration, but leaves 
behind a residual gauge invariance. It in turn corresponds to vanishing 
components of the current $\partial_\mu\tilde B^{a\mu\nu}$. (Technically,
there then is at least one flat direction in the otherwise Gaussian integrand.
The flat directions are along those eigenvectors of $\B$ possessing zero
eigenvalues.)

When incorporated with the exponent, which requires a regularisation \cite{Schaden:1989pz}, the determinant contributes a term
proportional to $\frac{1}{2}\ln\det\B$ to the action. This term together
with the $BB$ term constitutes the effective potential, which is obtained from
the exponent in the partition function after dropping all terms containing
derivatives of fields. The effective potential becomes singular for field 
configurations for which $\det\B=0$. It is gauge invariant because all
contributing addends are gauge invariant separately.

The classical equations of motion obtained by
varying the action in Eq.~(\ref{massless}) with respect to the dual 
antisymmetric tensor field $\tilde B^{a\mu\nu}$ are given by
\be
g^2\tilde B^a_{\mu\nu}
&=&
(g^\rho_\nu g^\sigma_\mu-g^\rho_\mu g^\sigma_\nu)
\partial_{\rho}(\B^{-1})^{ab}_{\sigma\kappa}(\partial_\lambda\tilde B^{b\lambda\kappa})
-
\nn
&&-
(\partial_\rho\tilde B^{d\rho\kappa})(\B^{-1})^{db}_{\kappa\mu}
f^{abc}(\B^{-1})^{ce}_{\nu\lambda}(\partial_\sigma\tilde B^{e\sigma\lambda}),
\nn
\ee
which coincides with the first of Eqs.~(\ref{cleoms}) with the field tensor
evaluated at the saddle point of the action, 
$F^a_{\mu\nu}[\asad(B)]$. Taking into account additionally the effect due to 
fluctuations of $A^a_\mu$ contributes a term proportional to 
$\sfrac{\delta\Det\B}{\delta\tilde B^{a\mu\nu}}\det^{-1}\B$ 
to the previous equation.

\subsection{Massive\label{FSRM}}

In the massive case the prototypical Lagrangian is of the form
${\cal L}={\cal L}_0+{\cal L}_m$, where
$
{\cal L}_m:=\sfrac{m^2}{2}A^a_\mu A^{a\mu}.
$
(Due to our conventions the physical mass is given by $m_\mathrm{phys}:=mg$.)
This contribution to the Lagrangian is of course not gauge invariant. Putting 
it, regardlessly, into the partition function, gives
\be
P
&=&
\int[dA][dU]
\exp\{i\sint d^4x[{\cal L}_0
+
\sfrac{m^2}{2}A^a_\mu A^{a\mu}]\},
\label{part3}
\ee
which can be interpreted as the unitary gauge representation of an extended
theory. In order to see this let us split the functional integral over 
$A^a_\mu$ into an integral over the gauge group $[dU]$ and gauge inequivalent 
field configurations $[dA]^\prime$. Usually this separation is carried out
by fixing a gauge according to
\be
\int[dA]^\prime
:=
\int[dA]\delta[f^a(A)-C^a]
\Delta_f(A).
\ee
$f^a(A)=C^a$ is the gauge condition and 
$\Delta_f(A)$ stands for
the Faddeev--Popov determinant
defined through $1\overset{!}{=}\int[dA]\delta[f^a(A)-C^a]\Delta_f(A)$
\footnote{Due to the possible occurrence of
what is known as Gribov copies \cite{Gribov:1977wm} this method might not achieve a unique
splitting of the two parts of the integration. For our illustrative purposes, 
however, this is not important.}. Introducing this reparametrisation into
the partition function (\ref{part3}) yields,
\be
P
&=&
\int[dA]^\prime[dU]\exp(i\sint d^4x\{
-\sfrac{1}{4g^2}F^a_{\mu\nu}F^{a\mu\nu}
+
\nn
&&+
\sfrac{m^2}{2}
[A_\mu-iU^\dagger(\partial_\mu U)]^a
[A^\mu-iU^\dagger(\partial^\mu U)]^a\}).
\label{party}
\nn
\ee
${\cal L}_0$ is gauge invariant in any case and remains thus
unaffected. In the mass term the gauge transformations appear explicitly
\cite{KG}.
We now replace all of these gauge transformations with an auxiliary
(gauge group valued) scalar field $\Phi$, $U^\dagger\rightarrow \Phi$, obeying the constraint
\be
\Phi^\dagger\Phi\overset{!}{\equiv}\mathbbm{1}.
\label{constraint}
\ee
The field $\Phi$ can be expressed as \mbox{$\Phi=:e^{-i\theta}$}, where
$\theta=:\theta^a T^a$ is the gauge algebra valued non-Abelian
generalisation of the St\"uckelberg field \cite{Stueckel}. For a massive
gauge theory they are a manifestation of the longitudinal degrees of freedom
of the gauge bosons. In the context of symmetry breaking they arise as
Goldstone modes ("pions"). In the context of the Thirring model these
observations have been made in \cite{Kondo:1996qa}. There it was noted as 
well that the $\theta$ is also the field used in the canonical Hamiltonian 
Batalin--Fradkin--Vilkovisky formalism \cite{BFV}. We can extract the 
manifestly gauge invariant classical Lagrangian
\be
{\cal L}_\mathrm{cl}
:=
-\sfrac{1}{4g^2}F^a_{\mu\nu}F^{a\mu\nu}
+m^2\tr[(D_\mu\Phi)^\dagger(D^\mu\Phi)],
\label{nlgsigma}
\ee
where $\Phi$ fields have been rearranged making use of the product rule of
differentiation and the cyclic property of the trace and where 
$D_\mu\Phi:=\partial_\mu\Phi-iA_\mu\Phi$. Eq.~(\ref{nlgsigma}) 
resembles the Lagrangian density of a non-linear gauged sigma model. 
In the Abelian case the fields $\theta$ decouple from the dynamics. For 
non-Abelian gauge groups they do not and one would have to deal with the
non-polynomial coupling to them. 

In the following we show that these spurious degrees of freedom can be
absorbed when making the transition to a formulation based on the
antisymmetric tensor field $B^a_{\mu\nu}$. Introducing the antisymmetric
tensor field into the corresponding partition function, like in the previous
section, results in,
\be
P
&\cong&
\int[dA][d\Phi][dB]\exp(i\sint d^4x
\times
\nn
&&\times
\{
-
\sfrac{g^2}{4}B^a_{\mu\nu}B^{a\mu\nu}
-
\sfrac{1}{2}\tilde F^a_{\mu\nu}B^{a\mu\nu}
+
\nn
&&~+
\sfrac{m^2}{2}
[A_\mu-i\Phi(\partial_\mu\Phi^\dagger)]^a
[A^\mu-i\Phi(\partial^\mu\Phi^\dagger)]^a\}).
\label{part3a}
\nn\ee
Removing the gauge scalars $\Phi$ from the mass term by a gauge
transformation of the gauge field $A^a_\mu$ makes them explicit in the $BF$
term,
\be
P
&=&
\int[dA][d\Phi][dB]\exp\{i\sint d^4x[
-
\sfrac{g^2}{4}B^a_{\mu\nu}B^{a\mu\nu}
-
\nn
&&-
\tr(\Phi\tilde F_{\mu\nu}\Phi^\dagger B^{\mu\nu})
+
\sfrac{m^2}{2}A^a_\mu A^{a\mu}]\}.
\ee
In the next step we would like to integrate over the Yang--Mills connection
$A^a_\mu$. Already in the previous expression, however, we can perceive that
the final result will only depend on the combination of fields $\Phi^\dagger
B_{\mu\nu}\Phi$. [The $\Phi$ field can also be made explicit in the $BB$ term
in form of the constraint (\ref{constraint}).] Therefore, the functional 
integral over $\Phi$ only covers multiple times the range which is already 
covered by the $[dB]$ integration. Hence the degrees of freedom of the field
$\Phi$ have become obsolete in this formulation and the $[d\Phi]$ integral 
can be factored out. Thus, we could have performed the unitary gauge 
calculation right from the start. In either case, the final result reads,
\be
P
&\cong&
\int[dB]\Det^{-\frac{1}{2}}\M\exp\{i\sint d^4x[
-
\sfrac{g^2}{4}B^a_{\mu\nu}B^{a\mu\nu}
-
\nn
&&-
\sfrac{1}{2}
(\partial_\kappa\tilde B^{a\kappa\mu})
(\M^{-1})^{ab}_{\mu\nu}
(\partial_\lambda\tilde B^{b\lambda\nu})
]\},
\label{part3b}
\ee
where 
$
\M^{ab}_{\mu\nu}:=\B^{ab}_{\mu\nu}-m^2\delta^{ab}g_{\mu\nu}
$,
which coincides with \cite{FT}.
$\M^{ab}_{\mu\nu}$ and hence $(\M^{-1})^{ab}_{\mu\nu}$ transform
homogeneously under the adjoint representation. In Eq.~(\ref{massless}) the
central matrix $(\B^{-1})^{ab}_{\mu\nu}$ in the analogous term transformed
in exactly the same way. There this behaviour ensured the gauge invariance
of this term's contribution to the classical action. Consequently, the 
classical action in the massive case has the same invariance properties. In
particular, the aforementioned gauge invariant classical action describes a
massive gauge theory without having to resort to additional scalar fields.
For $\det\B\neq 0$,
the limit $m\rightarrow 0$ is smooth. For $\det\B=0$ the conserved current
components alluded to above would have to be separated appropriately in order 
to recover the corresponding $\delta$ distributions present in these situations 
in the massless case.

Again the effective action is dominated by the term proportional to 
$\frac{1}{2}\det\M$. The contribution from the mass to $\M$ shifts the
eigenvalues from the values obtained for $\B$. Hence the singular
contributions are typically obtained for eigenvalues of $\B$ of the order of 
$m^2$. The effective potential is again gauge invariant, for the same reason
as in the massless case.

The classical equations of motion obtained by variation of the action in 
Eq.~(\ref{part3a}) are given by,
\be
g^2B^a_{\mu\nu}&=&-\tilde F^a_{\mu\nu},
\nn
D^{ab}_\mu(A)\tilde B^{b\mu\nu}
&=&
-m^2[A^\nu-i\Phi(\partial^\nu\Phi^\dagger)]^a,
\nn
0&=&\sfrac{\delta}{\delta\theta^b}
\sint d^4x\{[A_\mu-i\Phi(\partial_\mu\Phi^\dagger)]^a\}^2.
\label{cleommass}
\ee
In these equations a unique solution can be chosen, that is a gauge be fixed,
by selecting the scalar field $\Phi$. $\Phi\equiv 1$ gives the unitary
gauge, in which the last of the above equations drops out. 
The general non-Abelian case is difficult to handle already on the classical
level, which is one of the main motivations to look for an alternative
formulation. In the non-Abelian case, the equation of motion obtained from
Eq.~(\ref{part3b}) resembles strongly the massless case,
\be
g^2\tilde B^a_{\mu\nu}
&=&
(g^\rho_\nu g^\sigma_\mu-g^\rho_\mu g^\sigma_\nu)
\partial_{\rho}(\M^{-1})^{ab}_{\sigma\kappa}(\partial_\lambda\tilde B^{b\lambda\kappa})
-
\nn
&&-
(\partial_\rho\tilde B^{d\rho\kappa})(\M^{-1})^{db}_{\kappa\mu}
f^{abc}(\M^{-1})^{ce}_{\nu\lambda}(\partial_\sigma\tilde B^{e\sigma\lambda}),
\nonumber
\ee
insofar as all occurrences of $(\B^{-1})^{ab}_{\mu\nu}$ have been 
replaced by $(\M^{-1})^{ab}_{\mu\nu}$. Incorporation of the effect of the
Gaussian fluctuations of the gauge field $A_\mu^a$ would give rise to a
contribution proportional to 
$\sfrac{\delta\B}{\delta\tilde B^{a\mu\nu}}\det^{-1}\M$ in the 
previous equation.

Before we go over to more general cases of massive non-Abelian gauge field
theories, let us have a look at the weak coupling limit: There the $BB$ term in 
Eq.~(\ref{part3a}) is neglected. Subsequently, integrating out the 
$B^a_{\mu\nu}$ field enforces $F_{\mu\nu}^a\equiv 0$. [This condition also 
arises from the classical equation of motion (\ref{cleommass}) for $g$=0.] 
Hence, for vanishing coupling exclusively
pure gauge configurations of the gauge field $A^a_\mu$ contribute. They can
be combined with the $\Phi$ fields and one is left with a non-linear
realisation of a partition function,
\be
P\underset{(\ref{part3a})}{\overset{g=0}{\cong}}\int[d\Phi]
\exp\{im^2\sint d^4x~\tr[(\partial_\mu\Phi^\dagger)(\partial^\mu\Phi)]\},
\label{part3c}
\ee
of a free massless scalar \cite{FT}. Setting $g=0$ interchanges with
integrating out the $B^a_{\mu\nu}$ field from the partition function
(\ref{part3a}). Thus, the partition function (\ref{part3b}) with $g=0$ is 
equivalent to (\ref{part3c}). That a scalar degree of freedom can be described
by means of an antisymmetric tensor field has been noticed in \cite{KR}.

\subsubsection{Position-dependent mass and the Higgs\label{FSRH}}

One possible generalisation of the above set-up is obtained by softening
the constraint (\ref{constraint}). This can be seen as allowing for a
position dependent mass. The new degree of freedom ultimately corresponds to
the Higgs. When introducing the mass $m$ as new degree of freedom (as "mass
scalar") we can restrict its variation by introducing a potential term 
$V(m^2)$, which remains to be specified, and a kinetic term $K(m)$, which we 
choose in its canonical form $K(m)=\sfrac{1}{2}(\partial_\mu m)(\partial^\mu m)$. 
It gives a penalty for fast variations of $m$ between neighbouring space-time
points. The fixed mass model is obtained in the limit of an infinitely sharp
potential with its minimum located at a non-zero value for the mass. Putting 
together the partition function in unitary gauge leads to,
\be
P
&=&
\int[dA][dm]\exp\{i\sint d^4x[
-\sfrac{1}{4g^2}F^a_{\mu\nu}F^{a\mu\nu}
+
\nn
&&+
\sfrac{m^2}{2{\cal N}}A^a_\mu A^{a\mu}+K(m)+V(m^2)]\},
\label{part4}
\ee
where we have introduced the normalisation constant ${\cal
N}:=\mathrm{dim~R}$, with R standing for the representation of the scalars.
This factor allows us to keep the canonical normalisation of the mass scalar
$m$. We can now repeat the same steps as in the previous section in order to
identify the classical Lagrangian,
\be
{\cal L}_\mathrm{cl}
:=
-\sfrac{1}{4g^2}F^a_{\mu\nu}F^{a\mu\nu}
+{\cal N}^{-1}\tr[(D_\mu \phi)^\dagger(D^\mu \phi)]
+V(|\phi|^2),
\nonumber
\ee
where now $\phi:=m\Phi$. 
In order to reformulate the partition function in
terms of the antisymmetric tensor field we can once more repeat the steps in 
the previous section. Again the spurious degrees of freedom represented by 
the field $\Phi$ can be factored out. Finally, this gives \cite{SO},
\be
P
&\cong&
\int[dB][dm]\Det^{-\frac{1}{2}}\M\exp\{i\sint d^4x[
-
\sfrac{g^2}{4}B^a_{\mu\nu}B^{a\mu\nu}
-
\nn
&&-
\sfrac{1}{2}
(\partial_\kappa\tilde B^{a\kappa\mu})
(\M^{-1})^{ab}_{\mu\nu}
(\partial_\lambda\tilde B^{b\lambda\nu})
+
\nn
&&+
K(m)
+
V(m^2)
]\},
\label{part5}
\ee
where $\M^{ab}_{\mu\nu}=\B^{ab}_{\mu\nu}-m^2{\cal N}^{-1}\delta^{b}g_{\mu\nu}$ depends on the space-time dependent mass $m$.
The determinant can as usual be included with the exponent in form of a term
proportional to $\frac{1}{2}\det\M$, the pole of which will dominate
the effective potential. As just mentioned, however, $\M$ is also a function
of $m$. Hence, in order to find the minimum, the effective potential must
also be varied with respect to the mass $m$. 

Carrying the representation in terms of antisymmetric tensor fields another
step further, the partition function containing the kinetic term $K(m)$ of the 
mass scalar can be expressed as Abelian version of Eq.~(\ref{part3c}),
\be
&&\int[db][da]\exp\{i\sint d^4x
[-\sfrac{1}{2}\tilde b_{\mu\nu}f^{\mu\nu}+\sfrac{1}{2}a_\mu a^\mu]\}
=
\nn
&=&
\int[dm]\exp\{i\sint d^4x[\sfrac{1}{2}(\partial_\mu m)(\partial^\mu m)]\},
\ee
where here the mass scalar $m$ is identified with the Abelian gauge
parameter. Combining the last equation with the partition function
(\ref{part5}) all occurrences of the mass scalar $m$ can be replaced by the
phase integral $m\rightarrow\int dx^\mu a_\mu$. The $bf$ term enforces the
curvature $f$ to vanish which constrains $a_\mu$ to pure gauges
$\partial_\mu m$ and the aforementioned integral becomes path-independent.

\subsubsection{Non-diagonal mass term and the Weinberg--Salam model\label{NONDIAG}}

The mass terms investigated so far had in common that all the bosonic
degrees of freedom they described possessed the same mass. A more general mass term
would be given by
${\cal L}_m:=\sfrac{m^2}{2}A^a_\mu A^{b\mu}\m^{ab}$.
Another similar approach is based on the Lagrangian
${\cal L}_m:=\sfrac{m^2}{2}\tr\{A_\mu A^\mu\Psi\}$ where $\Psi$ is
group valued and constant. We shall begin our discussion with this second 
variant and limit
ourselves to a $\Psi$ with real entries and $\tr\Psi=1$, which, in fact, does 
not impose additional constraints. Using this expression in the 
partition function (\ref{part4}) and making explicit the gauge scalars yields,
\be
P
&=&
\int[dA][dm]\exp\{i\sint d^4x[
-\sfrac{1}{4g^2}F^a_{\mu\nu}F^{a\mu\nu}
+
\nn
&&+
\sfrac{1}{2}\tr\{(D_\mu\phi)^\dagger(D^\mu\phi)\Psi\}
+V(m^2)]\}.
\ee   
Expressed in terms of the antisymmetric tensor field $B^a_{\mu\nu}$, the
corresponding partition function coincides with Eq.~(\ref{part5}) but with
$\M^{ab}_{\mu\nu}$ replaced by 
$\M^{ab}_{\mu\nu}:=\B^{ab}_{\mu\nu}-m^2\tr\{T^aT^b\Psi\}g_{\mu\nu}$.

Let us now consider directly the \mbox{$SU(2)\times U(1)$} Weinberg--Salam model. Its
partition function can be expressed as,
\be
P
&=&
\int[dA][d\psi]\exp\{i\sint d^4x[-\sfrac{1}{4g^2}F^{\ul{a}}_{\mu\nu}F^{\ul{a}\mu\nu}
+
\nn
&&+
\sfrac{1}{2}
\psi^\dagger(\overleftarrow\partial_\mu+iA_\mu)
(\overrightarrow\partial^\mu-iA^\mu)\psi+V(|\psi|^2)]\},
\nonumber
\ee
where $\psi$ is a complex scalar doublet, $A_\mu:=A^{\ul{a}}_\mu T^{\ul{a}}$,
with $\ul{a}\in\{0;\dots;3\}$, $T^a$ here stands for the generators of $SU(2)$ in fundamental representation, and, 
accordingly, $T^0$ for $\sfrac{g_0}{2g}$ times the $2\times 2$ unit matrix, with
the $U(1)$ coupling constant $g_0$. The partition function can be 
reparametrised with $\psi=m\Phi\hat\psi$, where $m=\sqrt{|\psi|^2}$, $\Phi$ 
is a group valued scalar field as above, and  
$\hat\psi$ is a constant doublet with $|\hat\psi|^2=1$. The partition 
function then becomes,
\be
P
&=&
\int[dA][d\Phi][dm]\exp(i\sint d^4x\{
-\sfrac{1}{4g^2}F^{\ul{a}}_{\mu\nu}F^{\ul{a}\mu\nu}
+
\nn
&&+
\sfrac{m^2}{2}\tr[
\Phi^\dagger(\overleftarrow\partial_\mu+iA_\mu)
(\overrightarrow\partial^\mu-iA^\mu)\Phi\Psi]
+
\nn
&&+
\sfrac{1}{2}(\partial_\mu m)(\partial^\mu m)+V(m^2)\}),
\ee
where
\be
\Psi=\hat\psi\otimes\hat\psi^\dagger.
\label{Psi}
\ee
Making the transition to the first order formalism leads to
\be
P
&\cong&
\int[dA][dB][d\Phi][dm]\exp(i\sint d^4x\{
-\sfrac{g^2}{4}B^{\ul{a}}_{\mu\nu}B^{\ul{a}\mu\nu}
-
\nn
&&-\sfrac{1}{2}F^{\ul{a}}_{\mu\nu}\tilde B^{\ul{a}\mu\nu}
+
K(m)+V(m^2)
+
\nn
&&+
\sfrac{m^2}{2}\tr[
\Phi^\dagger(\overleftarrow\partial_\mu+iA_\mu)
(\overrightarrow\partial^\mu-iA^\mu)\Phi\Psi]\}).
\ee
As in the previous case, a gauge transformation of the gauge field $A_\mu^a$
can remove the gauge scalar $\Phi$ from the mass term (despite the matrix
$\Psi$). Thereafter $\Phi$ only appears in the combination
$\Phi^\dagger B_{\mu\nu}\Phi$ and the integral $[d\Phi]$ merely leads to
repetitions of the $[dB]$ integral. [The U(1) part drops out completely
right away.] Therefore the $[d\Phi]$ integration can be factored out,
\be
P
&\cong&
\int[dA][dB][dm]\exp\{i\sint d^4x[
-\sfrac{g^2}{4}B^{\ul{a}}_{\mu\nu}B^{\ul{a}\mu\nu}
-
\nn
&&-\sfrac{1}{2}F^{\ul{a}}_{\mu\nu}\tilde B^{\ul{a}\mu\nu}
+
\sfrac{m^2}{2}\tr(A_\mu A^\mu\Psi)
+
\nn
&&+
K(m)+V(m^2)]\}.
\ee
The subsequent integration over the gauge fields $A^{\ul{a}}_\mu$ leads to
\be
P
&\cong&
\int[dB][dm]\Det^{-\frac{1}{2}}\M\exp\{i\sint d^4x[
-
\sfrac{g^2}{4}B^{\ul{a}}_{\mu\nu}B^{\ul{a}\mu\nu}
-
\nn
&&-
\sfrac{1}{2}
(\partial_\kappa\tilde B^{\ul{a}\kappa\mu})
(\M^{-1})^{\ul{a}\ul{b}}_{\mu\nu}
(\partial_\lambda\tilde B^{\ul{b}\lambda\nu})
+
\nn
&&+
K(m)
+
V(m^2)
]\},
\label{part6}
\ee
where
$\M^{\ul{a}\ul{b}}_{\mu\nu}
:=
\B^{\ul{a}\ul{b}}_{\mu\nu}
-
m^2\tr(T^{\ul{a}}T^{\ul{b}}\Psi)g_{\mu\nu}$ and $\B^{\ul{a}\ul{b}}_{\mu\nu}=\B^{ab}_{mu\nu}~\forall~\ul{a},\ul{b}\neq 0$ and $\B^{\ul{a}\ul{b}}=0$ otherwise.

From hereon we continue our discussion based on the mass matrix
\be
\m^{\ul{a}\ul{b}}:=\sfrac{1}{2}\tr(\{T^{\ul{a}},T^{\ul{b}}\}\Psi),
\label{mama}
\ee
which had already been mentioned at the beginning of Sect.~\ref{NONDIAG}.
$\m^{\ul{a}\ul{b}}$ is real and has been chosen to be symmetric.
(Antisymmetric parts are projected out by the contraction with the symmetric
$A_\mu^{\ul{a}}A^{\ul{b}\mu}$.) Thus it possesses a complete orthonormal set 
of eigenvectors $\mu^{\ul{b}}_j$ with the associated real eigenvalues $m_j$, 
$\m^{\ul{a}\ul{b}}\mu^{\ul{b}}_j\overset{!}{=}{\not\hskip -0.9mm\Sigma}_jm_j\mu_j^{\ul{a}}$. With the
help of these normalised eigenvectors one can construct projectors 
$
\pi_j^{\ul{a}\ul{b}}
:=
{\not\hskip -0.9mm\Sigma}_j\mu_j^{\ul{a}}\mu_j^{\ul{b}}
$
and decompose the mass matrix, $\m^{\ul{a}\ul{b}}=m_j\pi_j^{\ul{a}\ul{b}}$.
The projectors are complete,
$\mathbbm{1}^{\ul{a}\ul{b}}=\Sigma_j\pi_j^{\ul{a}\ul{b}}$,
idempotent
${\not\hskip -0.9mm\Sigma}_j\pi_j^{\ul{a}\ul{b}}\pi_j^{\ul{b}\ul{c}}=\pi_j^{\ul{a}\ul{c}}$,
and satisfy
$\pi_j^{\ul{a}\ul{b}}\pi_k^{\ul{b}\ul{c}}\overset{j\neq k}{=}0$.
The matrix $\B^{\ul{a}\ul{b}}_{\mu\nu}$, the antisymmetric tensor field
$B^a_{\mu\nu}$, and the gauge field $A^a_\mu$ can also be decomposed with
the help of the eigenvectors:
$\B^{\ul{a}\ul{b}}_{\mu\nu}
=
\mu_j^{\ul{a}}\mathbbm{b}^{jk}_{\mu\nu}\mu_k^{\ul{b}}$,
where
$\mathbbm{b}^{jk}_{\mu\nu}
:=
\mu^{\ul{a}}_j\B^{\ul{a}\ul{b}}_{\mu\nu}\mu^{\ul{b}}_k$;
$B^{\ul{a}}_{\mu\nu}=b_{\mu\nu}^j\mu^{\ul{a}}_j$,
where
$b_{\mu\nu}^j:=B_{\mu\nu}^{\ul{a}}\mu^{\ul{a}}_j$;
and
$A^{\ul{a}}_\mu=a_\mu^j\mu^{\ul{a}}_j$,
where
$a_{\mu}^j:=A_{\mu}^{\ul{a}}\mu^{\ul{a}}_j$. Using this decomposition in the
partition function (\ref{part6}) leads to,
\be
P
&\cong&
\int[db][dm]\Det^{-\frac{1}{2}}\mathbbm{m}\exp\{i\sint d^4x[
-
\sfrac{g^2}{4}b^j_{\mu\nu}b^{j\mu\nu}
-
\nn
&&-
\sfrac{1}{2}
(\partial_\kappa\tilde b^{j\kappa\mu})
(\mathbbm{m}^{-1})^{jk}_{\mu\nu}
(\partial_\lambda\tilde b^{k\lambda\nu})
+
\nn
&&+
K(m)
+
V(m^2)
]\},
\label{part7}
\ee
where
$\mathbbm{m}^{jk}_{\mu\nu}
:=
\mathbbm{b}_{\mu\nu}^{jk}
-
m^2\sum_lm_l\delta^{jl}\delta^{kl}g_{\mu\nu}$.
Making use of the concrete form of $\m^{\ul{a}\ul{b}}$ given in
Eq.~(\ref{mama}), inserting $\Psi$ from Eq.~(\ref{Psi}), and subsequent
diagonalisation leads to the eigenvalues 0, $\sfrac{1}{4}$, $\sfrac{1}{4}$ and
\mbox{$\sfrac{1}{4}(1+\sfrac{g_0^2}{g^2})$}. These correspond to the photon,
the two W bosons and the heavier Z boson, respectively. 
The thus obtained tree-level Z to W mass ratio squared consistently reproduces the cosine of the 
Weinberg angle in terms of the coupling constants,
$\cos^2\vartheta_w=\frac{g^2}{g^2+g_0^2}$.
Due to the masslessness of the photon one addend in the sum over $l$ in the 
expression $\mathbbm{m}^{jk}_{\mu\nu}$ 
above does not contribute. Still, the total $\mathbbm{m}^{jk}_{\mu\nu}$ does
not vanish like in the case of a single massless Abelian gauge boson (see
Appendix \ref{appa}). Physically this corresponds to the coupling of the
photon to the W and Z bosons.

\subsubsection{Technicolour}

In technicolour \cite{TC} theories the standard model minus the Higgs is supplemented by
an additional strongly interacting sector containing fermions (techniquarks)
transforming under a given representation of the technicolour gauge group
and also charged under the electroweak gauge group. The electroweak symmetry
is broken by chiral symmetry breaking in the technicolour sector. From the point of view of modern collider experiments the most visible manifestations are those at relatively low energies, that is below the electroweak scale. A standard method to describe these signals is the construction of the corresponding low-energy theory. Its basic degrees of freedom are technicolour singlet fields like (pseudo)scalars and (axial) vectors. In connection with the present investigation we are most interested in the pseudoscalar sector and leave, for example the spin-one sector for a later study. The simplest breaking pattern of the flavour symmetry is $SU(2)_L\times SU(2)_R\rightarrow SU(2)_V$, which is realised for two techniflavours
transforming under a non-(pseudo)-real representation of the technicolour
group, gives rise to the three pions which become the longitudinal degrees
of freedom of the W and Z bosons. In this respect the model's low-energy
Lagrangian looks just like the standard model which has already been
discussed above. The breaking pattern becomes richer---and the number of pions
larger---either by increasing the number of flavours $N_f$ [$SU(N_f)_L\times
SU(N_f)_R\rightarrow SU(N_f)_V$] and/or in the presence of techniquarks
transforming under real [$SU(2N_f)\rightarrow SO(2N_f)$] or pseudreal
[$SU(2N_f)\rightarrow Sp(2N_f)$] representations; always assuming a breaking
down to the maximal diagonal subgroup. (For a survey of phenomenologically
viable technicolour models of these types see, e.g.
Ref.~\cite{Dietrich:2006cm}.) It turns out that the technicolour models
which are most favoured by electroweak precision data are walking
\cite{walk} (that is quasi-conformal) technicolour models which feature
techniquarks in higher dimensional representations of the gauge group. So
also the minimal walking technicolour model with two flavours in the adjoint
representation of $SU(2)$ \cite{Dietrich:2005jn}. The adjoint representation
is real which leads to the enhanced flavour symmetry $SU(4)$ which breaks to
$SO(4)$ yielding nine pions. Let us discuss this setup along the lines of
Ref.~\cite{Foadi:2007ue}. In the effective low-energy theory, the kinematic
term of these pions together with their scalar chiral partner provides the
mass term for the gauge bosons,
\be
{\cal L}_\mathrm{TC}
=
\sfrac{1}{2}\tr[({\cal D}_\mu M)({\cal D}^\mu M^\dagger)].
\label{ltc}
\ee
(We do not scale out the pion decay constant $f_\pi$ as in \cite{Foadi:2007ue}.)
Here $M$ transforms like the techniquark bilinear $M_{ij}\sim Q_i^\alpha Q_j^\beta\epsilon_{\alpha\beta}$ and in terms of low-energy fields can be parametrised according to 
\be
M=[\sfrac{1}{2}(\sigma+i\theta)+\sqrt{2}(i\Pi^a+\tilde{\Pi}^a)X^a]E.
\label{fields}
\ee
The matrix
\be
E:=\left(\begin{array}{cc}0&\mathbbm{1}\\\mathbbm{1}&0\end{array}\right)
\ee
characterises the condensate
$Q^\alpha_i Q^\beta_j\epsilon_{\alpha\beta}E^{ij}$
with the expectation value $2\langle M\rangle=vE$ in the basis
\be
Q:=\left(\begin{array}{r}U_L\\D_L\\-i\sigma^2U^*_R\\-i\sigma^2D^*_R\end{array}\right),
\ee
where $U_{L/R}$ and $D_{L/R}$ denote the left-/right-handed up and down techniquarks, respectively. The $X^a$ are the nine generators of $SU(4)$ which do not commute with the condensate $\sim E$. ${\cal D}_\mu$ stands for the electroweak covariant derivative. The electroweak group must be embedded in the $SU(4)$ in such a way that if the latter breaks to $SO(4)$ the former breaks from $SU(2)_L\times U(1)_Y$ to $U(1)_\mathrm{em}$. This is achieved by the choice $\sqrt{2}L^a:=S^a+X^a$ for the generators of $SU(2)_L$, $\sqrt{2}Y:=(S^3-X^3)^T+2Y_VS^4$ for the hypercharge generator and $\sqrt{2}Q=S^3=2Y_VS^4$.
$Y_V$ parametrises the hypercharge assignments for the techniquarks.
The $S^a$ are the four generators which leave the vacuum invariant, that is which commute with $E$. (For explicit expressions for the generators see Ref.~\cite{Appelquist:1999dq}.)
Then the covariant derivative reads,
\be
{\cal D}_\mu M
=
\partial_\mu M
-
i(G_\mu M+MG^T_\mu),
\ee
where
\be
G_\mu=A^a_\mu L^a+\sfrac{g_0}{g}A^0_\mu Y.
\ee

Due to the enhanced symmetry, the pion fields cannot be completely absorbed
by a gauge transformation of the gauge potential and the antisymmetric
tensor field. (This is also evident from another viewpoint; they carry
non-zero technibaryon number which clearly incompatible with the electroweak
gauge fields.) We therefore have,
\be
\sfrac{m^2}{2}{\bf m}^{\ul{a}\ul{b}}A^{\ul{a}}_\mu A^{\ul{b}\mu}
:=
\tr[(G_\mu M+MG^T_\mu)(G^\mu M+MG^{\mu T})^\dagger].
\nonumber
\ee
Further, let us define the current,
\be
mJ^{\ul{a}}_\mu A^{\ul{a}\mu}
&:=&
\sfrac{i}{2}\tr[(G_\mu M+MG^T_\mu)(-\partial_\mu M)^\dagger
+
\nn
&&+
(+\partial_\mu M)(G_\mu M+MG^T_\mu)^\dagger].
\ee
Structurally, the matrix ${\bf m}^{\ul{a}\ul{b}}$ has the same eigenvector
decomposition as explained after Eq.~(\ref{mama}). The radial degree of
freedom, denoted as above by $m^2$, is given by the sum over all the squares
of the fields present in Eq.~(\ref{fields}). In particular, it receives
contributions from the extra pion fields, which do not directly participate
in the breaking of the electroweak symmetry. Further, one of the eigenvalues
of $\mathbf{m}^{\ul{a}\ul{b}}$ is still zero, accounting for the massless photon. Also the current can be
decomposed in the eigenbasis of the mass matrix, $J^{\ul{a}}_\mu=J^j_\mu
\mu^{\ul{a}}_j$, where $J^j_\mu=J^{\ul{a}}_\mu\mu^{\ul{a}}_j$. The
Lagrangian density corresponding to the one in the exponent of
Eq.~(\ref{part7}) up to kinetic and potential terms for the (pseudo)scalars
then reads,
\be
{\cal L}_\mathrm{TC}
&\cong&
-
\sfrac{g^2}{4}b^j_{\mu\nu}b^{j\mu\nu}
-
\nn
&&-
\sfrac{1}{2}
(\partial_\kappa\tilde b^{j\kappa\mu}+mJ^{j\mu})
(\mathbbm{m}^{-1})^{jk}_{\mu\nu}
(\partial_\lambda\tilde b^{k\lambda\nu}+mJ^{k\nu})
\nonumber
\ee
where $\mathbbm{m}^{jk}_{\mu\nu}:=\mathbbm{b}^{jk}_{\mu\nu}-m^2\sum_l m_l\delta^{jl}\delta^{kl}g_{\mu\nu}$. The determinant induced by fluctuations of the gauge field is the same as in Eq.~(\ref{part7}).

Compared to the Weinberg--Salam model two types of additional terms arise
due to the presence of the current $J^{k\nu}$; on one hand, a
current-current interaction mediated by the restricted inverse mass matrix
$(\bar{\bf m}^{-1})^{\ul{a}\ul{b}}$; on the other, the current $J^{k\nu}$
acts as a source for the saddle point expression for the gauge field
$(\partial_\kappa\tilde
b^{j\kappa\mu}+mJ^{j\mu})(\mathbbm{m}^{-1})^{jk}_{\mu\nu}$.
 
We have here analysed the phenomenologically most preferred setting for
technicolour with an enhanced symmetry breaking pattern, two techniflavours
in the adjoint representation of $SU(2)$. The runner up, what viability is
concerned, is a model with two techniflavours in the two-index symmetric
representation of $SU(3)$. It possesses the simple breaking pattern which is
covered by the discussion of the Weinberg--Salam model. In
\cite{Dietrich:2006cm} other possibilities with larger symmetries are
listed. They will differ from the concrete example studied here by a
different number of extra pions. The main difference with respect to the
simple (Weinberg--Salam) patterns will, however, be the appearance of the 
extra momentum dependent current.

The inclusion of the corresponding technivector (technirho-s) and axial
vector fields would make the structure even richer. It is possible to
incorporate them in such a way that they can be second quantised (see for
example \cite{Foadi:2007ue}) which would allow us to treat them analogous to
the electroweak gauge bosons. This point, however, shall not be discussed here.

\section{Geometric representation\label{GEOM}}

The fact that the antisymmetric tensor field $B^a_{\mu\nu}$ transforms
homogeneously represents already an advantage over the formulation in terms
of the inhomogeneously transforming gauge fields $A^a_\mu$. Still, 
$B^a_{\mu\nu}$ contains degrees of freedom linked to the gauge transformations
(\ref{bigtrafo}). These
can be eliminated by making the transition to a formulation in terms of
geometric variables. 
In this section we provide a description of different
massive gauge field theories in terms of geometric variables in Euclidean 
space for two colours by adapting Ref.~\cite{Diakonov} to include mass.
The first-order action is quadratic in the gauge-field
$A^a_\mu$.\footnote{For two and/or three colours and four space-time 
dimensions there exist also other treatments of the massless setting 
\cite{lunev}, different from the one which here
is extended to the massive case.} Thus the evaluation of the classical
action at the saddle point yields the expression equivalent to the different 
exponents obtained after integrating out the gauge field $A^a_\mu$ in the
various partition functions in the previous section. 
In Euclidean space the classical massive Yang--Mills action in the first
order formalism reads
\be
S
:=
\int d^4x
({\cal L}_{BB}+{\cal L}_{BF}+{\cal L}_{AA}),
\ee
where
\be
{\cal L}_{BB}
&=&
-\sfrac{g^2}{4}B^a_{\mu\nu}B^a_{\mu\nu},
\label{BB}\\
{\cal L}_{BF}
&=&
+\sfrac{i}{4}\epsilon^{\mu\nu\kappa\lambda}B^a_{\mu\nu}F^a_{\kappa\lambda},
\label{BF}\\
{\cal L}_{AA}
&=&
-\sfrac{m^2}{2}A_\mu^a A^a_\mu.
\ee
At first we will investigate the
situation for the unitary gauge mass term ${\cal L}_{AA}$ and study the role
played by the scalars $\Phi$ afterwards.

As starting point it is important to note that a metric can
be constructed that makes the tensor $B^a_{\mu\nu}$ self-dual \cite{GS}. In
order to exploit this fact, it is convenient to define the antisymmetric 
tensor $(j\in\{1;2;3\})$
\be
T^j_{\mu\nu}:= \eta^j_{AB} e^A_\mu e^B_\nu,
\ee
with the self-dual 't Hooft symbol $\eta^j_{AB}$ \cite{tHooft}   
and the tetrad $e^A_\mu$. From there we construct a metric $\g_{\mu\nu}$ in 
terms of the tensor $T^j_{\mu\nu}$
\be
\g_{\mu\nu}
\equiv
e^A_\mu e^A_\nu
=
\sfrac{1}{6}\epsilon^{jkl}T^j_{\mu\kappa}T^{k\kappa\lambda}T^l_{\lambda\nu},
\ee
where
\be
T^{j\mu\nu}
:=
\sfrac{1}{2\sqrt{\g}}\epsilon^{\mu\nu\kappa\lambda}T^j_{\kappa\lambda}
\label{tinv}
\ee
and
\be
(\sqrt{\g})^3
&:=&
\sfrac{1}{48}
(\epsilon_{jkl}T^j_{\mu_1\nu_1}T^k_{\mu_2\nu_2}T^l_{\mu_3\nu_3})
\times
\nn
&&\times
(\epsilon_{j^\prime k^\prime l^\prime }
T^{j^\prime}_{\kappa_1\lambda_1}
T^{k^\prime}_{\kappa_2\lambda_2}
T^{l^\prime}_{\kappa_3\lambda_3})
\times
\nn
&&\times
\epsilon^{\mu_1\nu_1\kappa_1\lambda_1}
\epsilon^{\mu_2\nu_2\kappa_2\lambda_2}
\epsilon^{\mu_3\nu_3\kappa_3\lambda_3}
\ee
Subsequently, we introduce a triad $d_j^a$ such that
\be
B^a_{\mu\nu}=:d_j^a T^j_{\mu\nu}.
\label{bgeom}
\ee
This permits us to reexpress the $BB$ term of the classical Lagrangian,
\be
{\cal L}_{BB}
=
-\sfrac{g^2}{4}T^j_{\mu\nu}h_{jk}T^k_{\mu\nu},
\label{BBgeom}
\ee
where $h_{jk}:=d^a_jd^a_k$.
Putting Eqs.~(\ref{bgeom}) and (\ref{tinv}) into the saddle point
condition
\be
\sfrac{1}{2}\epsilon^{\kappa\lambda\mu\nu}D_\mu^{ab}(\asad)B^b_{\kappa\lambda}
=
+im^2\asad^a_\nu
\label{saddle}
\ee
gives
\be
D_\mu^{ab}(\asad)(\sqrt{\g}d^b_jT^{j\mu\nu})
=
+im^2\asad^a_\nu.
\ee
In the following we define the connection coefficients $\gamma_\mu|_j^k$ as
expansion parameters of the covariant derivative of the triads at
the saddle point in terms of the triads,
\be
D_\mu^{ab}(\asad)d^b_j=:\gamma_\mu|_j^k d^a_k.
\label{concoeff}
\ee
This would not be directly possible for more than two colours, as then
the set of triads is not complete. 
The connection coefficients allow us to define covariant derivatives 
according to
\be
\nabla_\mu|^k_j:=\partial_\mu\delta^k_j+\gamma_\mu|_j^k.
\label{covder}
\ee
These, in turn, permit us to rewrite the saddle point condition 
(\ref{saddle}) as
\be
d^a_k\nabla_\mu|^k_j(\sqrt{\g}T^{j\mu\nu})
=
im^2\asad^a_\nu,
\label{saddle2}
\ee
and the mass term in the classical Lagrangian becomes
\be
{\cal L}_{AA}
=
\sfrac{1}{2m^2}
[\nabla_\mu|^k_i(\sqrt{\g}T^{i\mu\nu})]
h_{kl}
[\nabla_\kappa|^l_j(\sqrt{\g}T^{j\kappa\nu})].
\label{massgeom}
\ee
In the limit \mbox{$m\rightarrow 0$} this term enforces the covariant conservation
condition $\nabla_\mu|^k_i(\sqrt{\g}T^{i\mu\nu})\equiv 0$, known for the 
massless case. It results also directly from the saddle point condition
(\ref{saddle2}). Here $d_k^a\nabla_\mu|^k_i(\sqrt{\g}T^{i\mu\nu})$ are the
direct analogues of the Abelian currents 
$\epsilon^{\mu\nu\kappa\lambda}\partial_\mu B_{\kappa\lambda}$, which are
conserved in the massless case [see Eq.~(\ref{abelpart1})] and distributed 
following a Gaussian distribution in the massive case [see
Eq.~(\ref{abelpart2})]. 

The commutator of the above covariant derivatives yields a Riemann-like 
tensor ${R^k}_{j\mu\nu}$ 
\be
{R^k}_{j\mu\nu}:=[\nabla_\mu,\nabla_\nu]^k_j.
\label{riemann}
\ee
By evaluating, in adjoint representation (marked by $\adj{~}$), the following difference of double
commutators 
$[\adj D_\mu(\asad),[\adj D_\nu(\asad),\adj d_j]]
-(\mu\leftrightarrow\nu)$
in two different ways, one can show that
\be
i[\adj d_j,\adj F_{\mu\nu}(\asad)]=\adj d_k {R^k}_{j\mu\nu},
\ee
or in components,
\be
F^a_{\mu\nu}(\asad)
=
\sfrac{1}{2}\epsilon^{abc}d^{bj}d^c_k{R^k}_{j\mu\nu},
\label{fgeom}
\ee
where $d^{aj}d^a_k:=\delta^j_k$ defines the inverse triad, 
$d^{aj}=h^{jk}d^a_k$. Hence,
we are now in the position to rewrite the remaining $BF$ term of the
Lagrangian density. Introducing Eqs.~(\ref{bgeom}) and (\ref{fgeom}) into 
Eq.~(\ref{BF}) results in
\be
{\cal L}_{BF}
=
\sfrac{i}{4}\sqrt{\g}
T^{j\mu\nu}{R^k}_{l\mu\nu}\epsilon_{jmk}h^{lm}.
\label{BFgeom}
\ee

Let us now repeat the previous steps with a mass
term in which the gauge scalars $\Phi$ are explicit,
\be
{\cal L}_{AA}^\Phi
:=
-\sfrac{m^2}{2}
[A_\mu-i\Phi(\partial_\mu\Phi^\dagger)]^a
[A_\mu-i\Phi(\partial_\mu\Phi^\dagger)]^a.
\ee
In that case the saddle point condition (\ref{saddle}) is given by,
\be
\sfrac{1}{2}\epsilon^{\kappa\lambda\mu\nu}
D_\mu^{ab}(\asad) B^b_{\kappa\lambda}
=
im^2
[\asad_\nu-i\Phi(\partial_\nu\Phi^\dagger)]^a,
\ee
or in the form of Eq.~(\ref{saddle2}), that is with the left-hand side
replaced,
\be
d^a_k\nabla_\mu|^k_j(\sqrt{\g}T^{j\mu\nu})
=
im^2[\asad_\nu-i\Phi(\partial_\nu\Phi^\dagger)]^a.
\ee
Reexpressing ${\cal L}_{AA}^\Phi$ with the help of the previous equation
reproduces exactly the unitary gauge result (\ref{massgeom}) for the mass 
term.

Finally, the tensor $\B$ appearing in the determinant (\ref{det}), which 
accounts for the Gaussian fluctuations of the gauge field $A^a_\mu$, 
formulated in the new variables reads 
$\B^{bc}_{\mu\nu}=\sqrt{\g}f^{abc}d^a_iT^{i\mu\nu}$.

The last ingredients required to put together the partition function is
the quantum measure in gauge invariant variables. Let us choose
$W_{\mu\nu\kappa\lambda}:=T^j_{\mu\nu}h_{jk}T^k_{\kappa\lambda}$. They are
antisymmetric in the first and the second pair of indices and symmetric
under exchange of the first pair with the second pair. In three
space-time dimensions they would suffice exactly to parametrise the six
gauge invariant degrees of freedom. The Jacobian required for the change of
variables would be,
\be
\int[dB]_3\cong\int[dW]_3\Det^{-\sfrac{1}{2}}J_3,
\ee
where
\be
J_3:=\left(\begin{array}{ccc}
W_{1212}&W_{1223}&W_{1231}\\
W_{2312}&W_{2323}&W_{2331}\\
W_{3112}&W_{3123}&W_{3131}
\end{array}\right).
\label{array}
\ee
In four space-time dimensions for $SU(2)$ not all $W_{\mu\nu\kappa\lambda}$ are independent
and we have to select a subset. One possible choice leads
to
\be
\int[dB]_4\cong\int[dW]_4\Det^{-\sfrac{1}{2}}J_4,
\label{W}
\ee
where
\be
J_4:=J_3(31)J_3(41)J_3(42)J_3(43),
\ee
the index pair ``31'' in Eq.~(\ref{array}) is each time replaced by the index pair in brackets, and the functional integral runs over all 15 components of
$W_{\mu\nu\kappa\lambda}$ contained in the Jacobian. 

For a
position-dependent mass the above discussion does not change materially. The
potential and kinematic term for the mass scalar $m$ have to be added
to the action.

Contrary to the massless case the $A_\mu^a$ dependent part of the Euclidean
action is genuinely complex. Without mass only the T-odd and hence purely
imaginary $BF$ term was $A_\mu^a$ dependent. With mass there contributes the
additional T-even and thus real mass term. Therefore the saddle point value
$\asad^a_\mu$ for the gauge field becomes complex. This is a known
phenomenon and in this context it is  essential to deform the integration
contour of the path integral in the partition function to run through the
saddle point \cite{CA}. For the Gaussian integrals which are under
consideration here, in doing so, we do not receive additional contributions. 
The imaginary part $\im\asad^a_\mu$ of the saddle point value of the gauge field transforms
homogeneously under gauge transformations. The complex valued saddle point
of the gauge field which is integrated out does not affect the
real-valuedness of the remaining fields, here $B^a_{\mu\nu}$. In this sense
the field $B^a_{\mu\nu}$ represents a parameter for the integration over
$A^a_\mu$. The tensor $T^j_{\mu\nu}$ is real-valued by definition and
therefore the same holds also for the triad $d^a_j$ [see Eq.~(\ref{bgeom})]. 
$h_{kl}$ is composed of the triads and, consequently, real-valued as well. 
The imaginary part of
the saddle point value of the gauge field, $\im\asad^a_\mu$, enters the
connection coefficients (\ref{concoeff}). Through them it affects the
covariant derivative (\ref{covder}) and the Riemann-like tensor
(\ref{riemann}). More concretely the connection coefficients
$\gamma_\mu|^k_j$ can be decomposed according to
\be
D^{ab}_\mu(\re\asad)d_j^b&=&(\re\gamma_\mu|^k_j)d_k^a,\nn
f^{abc}(\im\asad^c_\mu)d_j^b&=&(\im\gamma_\mu|^k_j)d_k^a,\nonumber
\ee
with the obvious consequences for the covariant derivative,
\be
\nabla_\mu|^k_j&=&\re\nabla_\mu|^k_j+i\im\nabla_\mu|^k_j,\nn
\re\nabla_\mu|^k_j&=&\partial_\mu\delta^k_j+\re\gamma_\mu|^k_j,\nn
\im\nabla_\mu|^k_j&=&\im\gamma_\mu|^k_j.\nonumber
\ee
This composition reflects in the mass term,
\be
\re{\cal L}_{AA}
&=&
\sfrac{1}{2m^2}\{[\re\nabla_\mu|^k_i(\sqrt{\g}T^{i\mu\nu})]h_{kl}
[\re\nabla_\kappa|^l_j(\sqrt{g}T^{j\kappa\nu})]
-
\nn
&&-
[\im\gamma_\mu|^k_j(\sqrt{\g}T^{i\mu\nu})]h_{kl}
[\im\gamma_\kappa|^l_j(\sqrt{\g}T^{j\kappa\nu})]\}\nn
\im{\cal L}_{AA}
&=&
\sfrac{2}{2m^2}[\re\nabla_\mu|^k_i(\sqrt{\g}T^{i\mu\nu})]h_{kl}
[\im\nabla_\kappa|^l_j(\sqrt{g}T^{j\kappa\nu})]\nonumber
\ee
on one hand, and in the Riemann-like tensor,
\be
\re {R^k}_{j\mu\nu}
&=&
[\re\nabla_\mu,\re\nabla_\nu]^k_j
-
[\im\nabla_\mu,\im\nabla_\nu]^k_j\nn
\im {R^k}_{j\mu\nu}
&=&
[\re\nabla_\mu,\im\nabla_\nu]^k_j
+
[\im\nabla_\mu,\re\nabla_\nu]^k_j.\nonumber
\ee
on the other. The connection to the imaginary part of $\asad^a_\mu$ is more
direct in Eq.~(\ref{fgeom}) which yields,
\be
\re F^a_{\mu\nu}(\asad)
&=&\sfrac{1}{2}\epsilon^{abc}d^{bj}d^c_k\re R^k{}_{j\mu\nu},\nn
\im F^a_{\mu\nu}(\asad)
&=&\sfrac{1}{2}\epsilon^{abc}d^{bj}d^c_k\im R^k{}_{j\mu\nu},\nonumber
\ee
Finally, the $BF$ term becomes,
\be
\re{\cal L}_{BF}
&=&
-\sfrac{1}{4}\sqrt{\g}T^{j\mu\nu}\epsilon_{jmk}h^{lm}\im R^k{}_{l\mu\nu},\nn
\im{\cal L}_{BF}
&=&
+\sfrac{1}{4}\sqrt{\g}T^{j\mu\nu}\epsilon_{jmk}h^{lm}\re R^k{}_{l\mu\nu}.\nonumber
\ee
Summing up, at the complex saddle point of the $[dA]$ integration the
emerging Euclidean ${\cal L}_{AA}$ and ${\cal L}_{BF}$ are both complex,
whereas before they were real and purely imaginary, respectively. Both terms
together determine the saddle point value $\asad^a_\mu$. Therefore, they
become coupled and cannot be considered separately any longer. This was already
to be expected from the analysis in Minkowski space in Sect.~\ref{FSR}, where
the matrix $\M^{ab}_{\mu\nu}$ combines T-odd and T-even contributions, which
originate from ${\cal L}_{AA}$ and ${\cal L}_{BF}$, respectively. There the
different contributions become entangled when the inverse
$(\M^{-1})^{ab}_{\mu\nu}$ is calculated.

\subsection{Weinberg--Salam model\label{GEOMWS}}

Now, let us reformulate the Weinberg--Salam model in geometric variables.
We omit here the kinematic term $K(m)$ and the potential term $V(m^2)$ for
the sake of brevity because they do not interfere with the calculations and 
can be reinstated at every time. The remaining terms of the classical action
are
\be
S&:=&\int d^4x(
{\cal L}^\mathrm{Abel}_{BB}+{\cal L}^\mathrm{Abel}_{BF}+
{\cal L}_{BB}+{\cal L}_{BF}+{\cal L}_{AA}),\nn
{\cal L}_{AA}
&:=&
-\sfrac{m^2}{2}\m^{\ul{a}\ul{b}}A^{\ul{a}}_\mu A^{\ul{b}}_{\mu},\label{laa}\\
{\cal L}^\mathrm{Abel}_{BB}&:=&-\sfrac{g^2}{4}B^0_{\mu\nu}B^{0}_{\mu\nu}
\label{AbelBB},\\
{\cal L}^\mathrm{Abel}_{BF}
&:=&
+\sfrac{i}{4}\epsilon^{\mu\nu\kappa\lambda}B^0_{\mu\nu}F^{0}_{\kappa\lambda},
\ee
and ${\cal L}_{BB}$ as well as ${\cal L}_{BF}$ have been defined in 
Eqs.~(\ref{BB}) and (\ref{BF}), respectively.

The saddle point conditions for the $[dA]$ integration with this action are
given by
\be
\sfrac{1}{2}\epsilon^{\kappa\lambda\mu\nu}D_\mu^{ab}(\asad)B^b_{\kappa\lambda}
&=&
+im^2\m^{a\ul{b}}A^{\ul{b}}_\nu,\\
\sfrac{1}{2}\epsilon^{\kappa\lambda\mu\nu}\partial_\mu B^0_{\kappa\lambda}
&=&
+im^2\m^{0\ul{b}}A^{\ul{b}}_\nu.
\ee
For the following it is convenient to use linear combinations of these
equations, which are obtained by contraction with the eigenvectors
$\mu^{\ul{a}}_l$ of the matrix $\m^{\ul{a}\ul{b}}$---defined between
Eqs.~(\ref{mama}) and (\ref{part7})---,
\be
\sfrac{1}{2}\epsilon^{\kappa\lambda\mu\nu}
[\mu^a_lD^{ab}_\mu(\asad)B^b_{\kappa\lambda}
+
\mu_l^0\partial_\mu B^0_{\kappa\lambda}]
=
im^2\mu_l^{\ul{a}}\m^{\ul{a}\ul{b}}A^{\ul{b}}_\nu.
\nn
\label{lincombs}
\ee
The non-Abelian term on the left-hand side can be rewritten using the
results from the first part of Sect.~\ref{GEOM}. The right-hand side may be
expressed in terms of eigenvalues of the matrix $\m^{\ul{a}\ul{b}}$. We find
(no summation over $l$),
\be
\mu_l^{\ul{a}}X^{\ul{a}\nu}=im^2m_la_\nu^l,
\label{X}
\ee
where
\be
X^{\ul{a}\nu}
:=
d_j^a\nabla_\mu|_k^j(\sqrt{\g}T^{k\mu\nu})
+
\sfrac{1}{2}\epsilon^{\kappa\lambda\mu\nu}
\mu_l^0\partial_\mu B_{\kappa\lambda}^0.
\label{x}
\ee
The mass term can be decomposed in the eigenbasis of $\m^{\ul{a}\ul{b}}$ as
well and, subsequently, be formulated in terms of the geometric variables,
\be
{\cal L}_{AA}
&=&
-\sfrac{m^2}{2}{\textstyle\sum}_lm_la_\mu^la_\mu^l
=
\nn
&=&
\sfrac{1}{2m^2}(\bar{\m}^{-1})^{\ul{a}\ul{b}}X^{\ul{a}\nu}X^{\ul{b}\nu},
\ee
where
\be
(\bar{\m}^{-1})^{\ul{a}\ul{b}}
:=
{\textstyle \sum}_l^{\forall m_l\neq 0}{m_l}^{-1}\mu_l^{\ul{a}}\mu^{\ul{b}}_l.
\label{minv}
\ee
Taking the mass eigenvalues to zero this addend leads to the covariant
conservation of the composite current 
in $X^{a\mu}\equiv 0$, just as previously observed in
the Abelian case in terms of gauge invariant antisymmetric tensor fields and
in the non-Abelian case with simple mass term in geometric variables. Like
in those situations for finite mass eigenvalues the magnitude of the 
aforementioned current components follow a Gaussian distribution. The mixture
of Abelian and non-Abelian currents is caused by the symmetry breaking pattern
$SU(2)_L\times U(1)_Y\rightarrow U(1)_\mathrm{em}$ which leaves unbroken
$U(1)_\mathrm{em}$ and not the $U(1)_Y$, which is a symmetry in the unbroken 
phase.

It should be emphasised that in the present geometric representation on the
classical level the introduction of a Higgs doublet does not inevitably
suggest itself, as its non-radial degrees of freedom are not needed to
ensure gauge invariance. In the quantised form the radial degree of freedom
of the Higgs takes of course care of perturbative renormalisability.

With the help of the above relations and the results from the beginning of
Sect.~\ref{GEOM} we are now in the position to express the classical
action in geometric variables: The mass term is given in the previous
expression. 
The Abelian antisymmetric fields
$B^0_{\mu\nu}$ in ${\cal L}^\mathrm{Abel}_{BB}$ are gauge invariant and we
leave ${\cal L}^\mathrm{Abel}_{BB}$ as defined in Eq.~(\ref{AbelBB}). In
geometric variables ${\cal L}_{BB}$ is given by Eq.~(\ref{BBgeom}) and 
${\cal L}_{BF}$ by Eq.~(\ref{BFgeom}). 
At the end the kinetic term $K(m)$ and the potential term $V(m^2)$ should be 
reinstated. 

Additional contributions from fluctuations give rise to an addend (on
the level of the Lagrangian) proportional to 
$\sfrac{1}{2}\ln\det\mathbbm{m}$, where $\mathbbm{m}$ 
can be expressed in the new variables, 
$\mathbbm{m}^{jk}_{\mu\nu}
=
f^{abc}d^a_l\mu_j^b\mu^c_k\sqrt{\g}T^{l\mu\nu}
-
m^2\sum_lm_l\delta^{lj}\delta^{kl}g_{\mu\nu}$.

In order to reconstruct the partition function only the measure of the
functional integral has to be translated into gauge invariant variables. The
Abelian antisymmetric field $B^0_{\mu\nu}$ is already gauge invariant and
can be kept as variable. The integral over the non-Abelian fields can be 
reexpressed like in Eq.~(\ref{W}).

Repeating the entire calculation not in unitary gauge, but with explicit gauge
scalars $\Phi$, yields exactly the same result because the mass term and the
saddle point condition change in unison, such that Eq.~(\ref{X}) is obtained
again. This has already been demonstrated explicitly for a massive Yang--Mills
theory just before Sect.~\ref{GEOMWS}.

\subsection{Technicolour\label{GEOMTC}}

We begin by replacing Eq.~(\ref{laa}) by Eq.~(\ref{ltc}). Then the saddle point conditions for the variation of the corresponding classical action with respect to the gauge potentials $A^{\ul{a}}_\mu$ reads,
\be
\sfrac{1}{2}\epsilon^{\kappa\lambda\mu\nu}D_\mu^{ab}(\asad)B^b_{\kappa\lambda}
&=&
+im^2\m^{a\ul{b}}\asad^{\ul{b}}_\nu+mJ^a_\nu,\\
\sfrac{1}{2}\epsilon^{\kappa\lambda\mu\nu}\partial_\mu B^0_{\kappa\lambda}
&=&
+im^2\m^{0\ul{b}}\asad^{\ul{b}}_\nu+mJ^0_\nu.
\label{saddletc}
\ee
The linear combinations analogous to Eq.~(\ref{lincombs}) are given by,
\be
\sfrac{1}{2}\epsilon^{\kappa\lambda\mu\nu}
[\mu^a_lD^{ab}_\mu(\asad)B^b_{\kappa\lambda}
+
\mu_l^0\partial_\mu B^0_{\kappa\lambda}]
=
\mu_l^{\ul{a}}(im^2\m^{\ul{a}\ul{b}}A^{\ul{b}}_\nu+mJ^{\ul{a}}_\nu).
\nonumber
\ee
Reexpressing the covariant derivative on the left-hand side in terms of
gauge singlet variables as described in Sect.~\ref{GEOM} we obtain (no
summation over $l$),
\be
\mu_l^{\ul{a}}(X^{\ul{a}\nu}-mJ^{\ul{a}\nu})=im^2m_la_\nu^l,
\label{XTC}
\ee
where $X^{\ul{a}\nu}$ has been defined in Eq.~(\ref{x}). Evaluating the Lagrangian at the saddle point for the gauge potential we find,
\be
{\cal L}_\mathrm{AA}+{\cal L}_\mathrm{JA}
&=&
-\sfrac{m^2}{2}{\textstyle\sum}_l m_l a_\mu^l a_\mu^l
+
imJ^l_\mu a^l_\mu
\rightarrow
\nn
&\rightarrow&
\sfrac{1}{2m^2}(\bar{\m}^{-1})^{\ul{a}\ul{b}}
(X^{\ul{a}\nu}X^{\ul{b}\nu}-m^2J^{\ul{a}\nu}J^{\ul{b}\nu}),
\nonumber
\ee
with $(\bar{\m}^{-1})^{\ul{a}\ul{b}}$ from Eq.~(\ref{minv}).
Fluctuations of the gauge field induce the determinant term which here on
the Lagrangian level is proportional to $\sfrac{1}{2}\ln\det\mathbbm{m}$,
where again
$\mathbbm{m}^{jk}_{\mu\nu}:=f^{abc}d_l^a\mu^b_j\mu^c_k\sqrt{\g}T^{l\mu\nu}-m^2\sum_lm_l\delta^{lj}\delta^{kl}g_{\mu\nu}$.
As gauge invariant quantum measure for the gauge field sector we can use the
expression from Eq.~(\ref{W}).

Hence, outwardly it at first looks as if, apart from the combination in the
radial degree of freedom $m^2$, the degrees of freedom in the current
$X^{\ul{a}\nu}$ decouple from those in the current $J^{\ul{a}\nu}$ at tree
level, as long as we do not take into account potential terms for the $M$
field. This is, however, not the case because the current $J^{\ul{a}\nu}$
enters the connection coefficient $\gamma_\mu |^j_k$ through the saddle
point condition (\ref{saddletc}) and from there feeds into the current
$X^{\ul{a}\nu}$ and the tensor $R^k{}_{l\mu\nu}$. This corresponds to the
coupling of said current to the saddle point expression of the vector
potential alluded to before in the representation based on antisymmetric
tensor fields. The self interaction between the currents is the same as in
that case.

\subsection{Link to torsion\label{TOR}}

Here we interpret the above result by pointing out that in three space-time
dimensions a non-zero mass leads to the presence of torsion in the
geometrical description of SU(2) gauge groups. In three dimensions we
introduce a one-form $E$ instead of the two-form $B$
because the dual of the two-form will be the required two-form. Thus, the
Lagrangian density reads,
\be
{\cal L}:={\cal L}_{EE}+{\cal L}_{EF}+{\cal L}_{AA},
\ee
where
\be
{\cal L}_{EE}&:=&-\sfrac{g^2}{2}E^a_\mu E^a_\mu ,\\
{\cal L}_{EF}
&:=&
+\sfrac{i}{2}\epsilon^{\mu\nu\lambda}E^a_\mu F^a_{\nu\lambda} ,\\
{\cal L}_{AA}&:=&-\sfrac{m^2}{2}A^a_\mu A^a_\mu .
\ee
We stick to a simple diagonal mass term which is sufficient to understand
the reasoning. As next step, we reexpress the mass term through the introduction of a
Lagrange multiplier field $C^a_\mu$,
\be
{\cal L}_{AA}\rightarrow -\sfrac{1}{2}C^a_\mu C^a_\mu+imC^a_\mu A^a_\mu .
\ee
For the action constructed from the resulting total Lagrangian the saddle 
point condition reads,
\be
\epsilon^{\mu\nu\lambda}D^{ab}_\mu(\asad)E^b_\nu=mC^a_\lambda.
\label{saddlec}
\ee
In three dimensions we can interpret the $E^a_\nu$ directly as, in general,
complete set of dreibeinen and define the connection coefficients 
$\Gamma_\mu|^\kappa_\nu$,
\be
D^{ab}_\mu(\asad)E^b_\nu=:\Gamma_\mu|^\kappa_\nu E^a_\kappa.
\ee
The general solution of Eq.~(\ref{saddlec}) for vanishing mass $m$ is given 
by symmetric connection coefficients,
\be
\epsilon^{\mu\nu\lambda}\Gamma_\mu|^\kappa_\nu=0.
\ee
For finite mass $m$ $C^a_\mu$ is equal to the antisymmetric part of the 
connection coefficient,
\be
\epsilon^{\mu\nu\lambda}\Gamma_\mu|^\kappa_\nu E^a_\kappa=mC^a_\lambda,
\label{torsionc}
\ee
that is the torsion. The relation obtained by variation with respect to 
$C^a_\mu$ reads,
\be
C^a_\mu=im\asad^a_\mu.
\ee
Combined with Eq.~(\ref{torsionc}) this yields,
\be
\epsilon^{\mu\nu\lambda}\Gamma_\mu|^\kappa_\nu E^a_\kappa=im^2\asad^a_\lambda,
\ee
which is the equivalent of Eq.~(\ref{saddle2}).

Adding another term to the total Lagrangian density, ${\cal
L}\rightarrow{\cal L}+{\cal L}_{JA}$, like the one arising from the extra 
pions in technicolour with enhanced symmetry, 
\be
{\cal L}_{JA}:=imJ^a_\mu A^a_\mu,
\ee 
yields another contribution to the torsion,
\be
\epsilon^{\mu\nu\lambda}\Gamma_\mu|^\kappa_\nu E^a_\kappa
=
mC^a_\lambda+mJ^a_\lambda.
\ee
The two contributions differ insofar as that $C^a_\mu$ does not contain
derivatives of the underlying fields and $J^a_\mu$ is exclusively first order 
in derivatives. Therefore, at very low energies $J^a_\mu$ is suppressed 
relative to $C^a_\mu$. Hence, there the $C^a_\mu$ contribution to torsion 
dominates. At higher scales the momentum dependent contribution from $J^a_\mu$ 
becomes increasingly important. 

For electroweak symmetry breaking the mass matrix 
${\bf m}^{\underline{a}\underline{b}}$ or its restricted inverse 
$(\bar{\bf m}^{-1})^{\underline{a}\underline{b}}$ have to be reinstated
together with the contributions from the hypercharge field. Neither of these
two steps, however, does fundamentally alter what was just said. Going to
four space-time dimensions necessitates a generalisation of the concept of
torsion. This is most directly perceptible by comparing the connection
coefficients $\Gamma_\mu|^\kappa_\nu$ and $\gamma_\mu|^k_j$. The latter
feature a mismatch between the dimensions of the two lower indices, which
inhibits the standard definition of torsion. What for four space-time
dimensions replaces the antisymmetric part of $\Gamma_\mu|^\kappa_\nu$ is the 
current $X^{\underline{a}\nu}$ given in Eq.~(\ref{x}).

\section{Summary\label{SUM}}

We have here derived manifestly gauge-invariant formulations for theories 
breaking the electroweak symmetry. Namely, we have studied the
standard model case, that is the Weinberg--Salam model, and dynamical
electroweak symmetry breaking through technicolour models. For each approach the
derivation proceeded through two stages. The outset was always the standard
formulation in terms of Yang--Mills gauge potentials. This field transforms
inhomogeneously under gauge transformations. From there we
introduced antisymmetric tensor fields and subsequently eliminated the
Yang--Mills potential. The antisymmetric tensor fields transform
ultralocally under gauge transformation. In this way St\"uckelberg degrees
of freedom, which were required for a gauge invariant formulation of a
massive gauge field theory in terms of Yang--Mills potentials, become obsolete. Still, the antisymmetric tensor
fields are no gauge singlets. Therefore, in the second stage we have introduced
gauge singlet variables, which lead to a formulation in terms of geometric
quantities. In this framework we have linked the presence of massive gauge
bosons to the presence of torsion in the geometric representation. More
precisely, in three space-time dimensions and for a three-dimensional gauge
group like the relevant $SU(2)_L$ a massless theory corresponds to a
torsionless geometric description. When mass is included the torsion becomes
non vanishing, but follows a Gaussian distribution centred around zero and 
the width of which is given by the mass. For other combinations of the
number of space-time dimensions and the dimension of the gauge group a
generalisation of the concept of torsion is necessary.

The mass-generation for the gauge bosons of the electroweak interactions 
possesses a number of non-standard features: It exhibits a non-diagonal 
breaking pattern, $SU(2)_L\times U(1)_Y\rightarrow U(1)_\mathrm{em}$, and the 
related non-diagonal mass term with, on top, a zero eigenvalue for the
massless photon, on one hand, and the position-dependent mass,
that is the Higgs degree of freedom needed for perturbative
renormalisability. In order to disentangle which characteristic of the
alternative formulations arises from which trait of the massive and
non-Abelian gauge theory, we first present the translation for a massless
non-Abelian theory. We then continue with massive non-Abelian theories with
diagonal mass terms and constant mass. The next generalisation is mandated
by the requirement of perturbative renormalisability and leads to a
position-dependent mass also known as Higgs degree of freedom. In order to
be able to accommodate the phenomenologically relevant breaking pattern the
generalisation to a non-diagonal mass term has to be performed. What is
referred to as the Higgs doublet is a combination of this radial degree of 
freedom and the aforementioned St\"uckelberg degree of freedom. In the
treatment with Yang--Mills potential one commonly picks an "expectation
value" for the Higgs doublet field in order to obtain a convenient 
parametrisation. The notion of an "expectation value" for the Higgs doublet
field is, however, misleading as in reality it is not even necessarily 
different from zero \cite{Frohlich:1980gj}. This discord can also be avoided
in the standard formulation. In the alternative formulations spelled out
here, though, this is automatic. 

Technicolour models which pass the constraints from currently available
electroweak precision data have, in general, larger flavour symmetries than
the minimally necessary $SU(2)_L\times SU(2)_R$ which by breaking to
$SU(2)_V$ provides the three necessary longitudinal degrees of freedom for
the W and Z bosons. For this reason, they have a richer
low-energy particle content, among which are the additional pseudoscalars
of importance to this investigation. The extras do not correspond to
St\"uckelberg degrees of freedom of the electroweak gauge symmetry and can
accordingly not be absorbed in the antisymmetric tensor fields and also
appear explicitly in the geometric formulation. In the $B$-field
formulation currents constructed from these additional fields act as source
for the saddle point expression of the vector potential expressed in terms
of the B-fields. There, and later on in the geometric representation a term
quadratic in the currents appears. It is of fourth order in the fields and
of second order in derivatives. In general, because said current contains
one derivative, it decouples at small momenta. In the geometric description
it enters in the definition of the connection coefficients and from there
the Riemann-like tensor.

\section*{Acknowledgments}

The author would like to thank
Luigi Del Debbio,
Gerald Dunne,
Roshan Foadi,
Mads T.~Frandsen,
Stefan Hofmann,
Matti J\"arvinen,
Thomas B.~Madsen, 
Francesco Sannino,
Martin Svensson,
and
Andrew Swann
for discussions.
Thanks are again due to Francesco Sannino for a thorough reading of and useful comments on the manuscript.  
The work was supported by the Danish Natural Science Research Council.

\appendix
\section{Abelian\label{appa}}
\subsection{Massless}

The partition function of an Abelian gauge field theory without fermions is
given by
\be
P:=\int[dA]\exp\{i\sint d^4x{\cal L}\}
\ee
with the Lagrangian density
\be
{\cal L}={\cal L}_0:=-\sfrac{1}{4g^2}F_{\mu\nu}F^{\mu\nu}
\label{l0}
\ee
and the field tensor
\be
F_{\mu\nu}:=\partial_\mu A_\nu-\partial_\nu A_\mu.
\ee
$g$ stands for the coupling constant. The transition to the
first-order formalism can be performed just like in the non-Abelian case,
which is treated in the main body of the paper. We find the partition function,
\be
P
&=&
\int[dA][dB]
\times
\nn
&&\times
\exp\{i\sint d^4x[
-\sfrac{1}{2}\tilde F_{\mu\nu}B^{\mu\nu}
-\sfrac{g^2}{4}B_{\mu\nu}B^{\mu\nu}]\}.
\nonumber
\ee
Here the antisymmetric tensor field $B_{\mu\nu}$, like the field tensor
$F_{\mu\nu}$, is gauge invariant. The classical equations of motion are
given by
\be
\partial_\mu\tilde B^{\mu\nu}=0~~~\mathrm{and}~~~g^2B_{\mu\nu}=-\tilde
F_{\mu\nu},
\ee
which after elimination of $B_{\mu\nu}$ reproduce the Maxwell equations one
would obtain from Eq.~(\ref{l0}).
Now we can formally integrate out the gauge field $A_\mu$. As no gauge is
fixed by the $BF$ term because the Abelian field tensor $F_{\mu\nu}$ is gauge
invariant this gives rise to a functional $\delta$ distribution. This 
constrains the allowed field configurations to those for which the conserved 
current $\partial_\mu\tilde B^{\mu\nu}$ vanishes,
\be
P
&\cong&
\int[dB]
\delta(\partial_\mu\tilde B^{\mu\nu})
\exp\{i\sint d^4x[-\sfrac{g^2}{4}B_{\mu\nu}B^{\mu\nu}]\}.
\nn
\label{abelpart1}
\ee

\subsection{Massive}

In the massive case the Lagrangian density becomes
${\cal L}={\cal L}_0+{\cal L}_m$, where
${\cal L}_m:=\sfrac{m^2}{2}A_\mu A^\mu$.
First, we here repeat some steps carried out above in the non-Abelian case: 
We can directly write down the partition function in unitary gauge. Regauging 
like in Eq.~(\ref{party}) leads to
\be
P
&=&
\int[dA]^\prime[dU]\exp(i\sint d^4x\{
-\sfrac{1}{4g^2}F_{\mu\nu}F^{\mu\nu}
+
\nn
&&+
\sfrac{m^2}{2}
[A_\mu-iU^\dagger(\partial_\mu U)]
[A^\mu-iU^\dagger(\partial^\mu U)]\}).
\nn
\label{nlinsig}
\ee
The corresponding gauge-invariant Lagrangian then reads,
\be
{\cal L}_\mathrm{cl}
:=
-\sfrac{1}{4g^2}F_{\mu\nu}F^{\mu\nu}+\sfrac{m^2}{2}(D_\mu\Phi)^\dagger(D^\mu\Phi),
\ee
with the constraint $\Phi^\dagger\Phi\overset{!}{=}1$. Constructing a
partition function in the first-order formalism from the previous Lagrangian 
yields,
\be
P
&\cong&
\int[dA][d\Phi][dB]
\times
\nn
&&\times
\exp(i\sint d^4x\{
-
\sfrac{1}{2}B_{\mu\nu}\tilde F^{\mu\nu}
-
\sfrac{g^2}{4}B_{\mu\nu}B^{\mu\nu}
+
\nn
&&+
\sfrac{m^2}{2}
[A_\mu-i\Phi(\partial_\mu\Phi^\dagger)]
[A^\mu-i\Phi(\partial^\mu\Phi^\dagger)]\}).
\nonumber
\ee
The $\Phi$ fields can be absorbed entirely in a gauge-transformation of the
gauge field $A_\mu$. The integration over $\Phi$ decouples. This can also be
seen by putting the parametrisation $\Phi=e^{-i\theta}$ into the previous 
equation and carrying out the $[dA]$ integration,  
\be
P
&\cong&
\int[dB][d\theta]\exp\{i\sint d^4x [
-\sfrac{g^2}{4}B_{\mu\nu}B^{\mu\nu}
-
\nn
&&-
\sfrac{1}{2m^2}
(\partial_\kappa\tilde B^{\kappa\mu})g_{\mu\nu}
(\partial_\lambda\tilde B^{\lambda\nu})
-
(\partial_\mu\theta)(\partial_\kappa B^{\kappa\mu})
]\}.
\label{abelpart2}
\nn
\ee
The only $\theta$ dependent term in the exponent is a total derivative and
drops out, leading to a factorisation of the $\theta$ integral.
Contrary to the non-Abelian case there arises no fluctuation
determinant depending on dynamical fields. Hence, in the Abelian case
starting with the classical action in Eq.~(\ref{abelpart2}) 
or the non-linear sigma model (\ref{nlinsig}) is equivalent 
\cite{Harikumar:1996sf}.

A third way which yields the same final result, starts by integrating out the
$\theta$ field first. This gives a transverse mass term
$\sim A^\mu(g_{\mu\nu}-\sfrac{\partial_\mu\partial_\nu}{\Box})A^\nu$.
Integration over $A_\mu$ then leads to the same result as before.

Instead of a vanishing current $\partial_\mu\tilde B^{\mu\nu}$ like in the
massless case, in the massive case the current has a Gaussian distribution.
The distribution's width is proportional to the mass of the gauge boson.

\subsubsection*{$m\rightarrow 0$ limit}

In the gauge-field representation the massless limit for the classical
actions discussed above are smooth. In terms of the $B_{\mu\nu}$ field the
mass $m$ ends up in the denominator of the corresponding term in the action. 
Together with the $m$ dependent
normalisation factors arising form the integrations over the gauge-field in
the course of the derivation of the $B_{\mu\nu}$ representation, however,
the limit $m\rightarrow 0$ still yields the $m=0$ result for the partition
function (\ref{abelpart1}). 

Still, it is known that the perturbative propagator for a massive photon 
is ill-defined if the mass goes to zero:
Variation of the exponent of the Abelian massive partition function in
unitary gauge with
respect to $A_\kappa$ and $A_\lambda$ gives the inverse propagator for the
gauge fields,
\be
(G^{-1})^{\kappa\lambda}
=
[(p^2-m^2_\mathrm{phys})g^{\kappa\lambda}-p^\kappa p^\lambda],
\ee
which here is already transformed to momentum space. The corresponding
equation of motion,
\be
(G^{-1})^{\kappa\lambda}
G_{\lambda\mu}\overset{!}{=}g^\kappa_\mu,
\ee
is solved by
\be
G_{\lambda\mu}
=
\frac{g_{\lambda\mu}}{p^2-m^2_\mathrm{phys}}
-
\frac{1}{m^2_\mathrm{phys}}\frac{p_\lambda p_\mu}{p^2-m^2_\mathrm{phys}},
\ee
with boundary conditions (an $\epsilon$ prescription) to be specified and
$m_\mathrm{phys}:=mg$. This propagator diverges in the limit $m\rightarrow 0$. 

In the representation based on the antisymmetric tensor fields, variation of
the exponent of the partition function (\ref{abelpart2}) with respect to the
fields $\tilde B_{\mu\nu}$ and $\tilde B_{\kappa\lambda}$ yields the inverse
propagator
\be
(G^{-1})^{\mu\nu|\kappa\lambda}
&=&
g^{\mu\kappa}g^{\nu\lambda}-g^{\nu\kappa}g^{\mu\lambda}
+
\nn
&&+
m^{-2}_\mathrm{phys}
(\partial^\mu\partial^\kappa g^{\nu\lambda}
-\partial^\nu\partial^\kappa g^{\mu\lambda}
-
\nn
&&~-
\partial^\mu\partial^\lambda g^{\nu\kappa}
+\partial^\nu\partial^\lambda g^{\mu\kappa}),
\ee
already expressed in momentum space. Variation with respect to $\tilde
B_{\mu\nu}$ instead of $B_{\mu\nu}$ corresponds only to a reshuffling of the
Lorentz indices and gives an equivalent description. The antisymmetric
structure of the inverse propagator is due to the antisymmetry of $\tilde
B_{\mu\nu}$. The equation of motion is then given by
\be
(G^{-1})^{\mu\nu|\kappa\lambda}G_{\kappa\lambda|\rho\sigma}
\overset{!}{=}
g^\mu_\rho g^\nu_\sigma-g^\mu_\sigma g^\nu_\rho
\ee
and solved by
\be
2G_{\kappa\lambda|\rho\sigma}
&=&
(g_{\kappa\rho}g_{\lambda\sigma}-g_{\kappa\sigma}g_{\lambda\rho})
-
\frac{1}{p^2-m^2_\mathrm{phys}}
\times
\nn
&&\times
(p_\kappa p_\rho g_{\lambda\sigma}
-p_\kappa p_\sigma g_{\lambda\rho}
-
\nn
&&~
p_\lambda p_\rho g_{\kappa\sigma}
+p_\lambda p_\sigma g_{\kappa\rho}).
\ee
Here we observe that the limit $m\rightarrow 0$ is well-defined,
\be
2G_{\kappa\lambda|\rho\sigma}
&\xrightarrow{m\rightarrow 0}&
g_{\kappa\rho}g_{\lambda\sigma}-g_{\kappa\sigma}g_{\lambda\rho}
-
\nn
&&-
\frac{1}{p^2}
(p_\kappa p_\rho g_{\lambda\sigma}
-p_\kappa p_\sigma g_{\lambda\rho}
-
\nn
&&~-
p_\lambda p_\rho g_{\kappa\sigma}
+p_\lambda p_\sigma g_{\kappa\rho}).
\ee
This is due to the consistent treatment of the gauge degrees of freedom
in the second approach.



\begin{thebibliography}{99}


\bibitem{Wilson}
A.~M.~Polyakov,
  Nucl.\ Phys.\  B {\bf 164} (1980) 171;\\
Yu.~M.~Makeenko and A.~A.~Migdal,
  Phys.\ Lett.\  B {\bf 88} (1979) 135
  [Erratum-ibid.\  B {\bf 89} (1980) 437];\\
  Nucl.\ Phys.\  B {\bf 188} (1981) 269
  [Sov.\ J.\ Nucl.\ Phys.\  {\bf 32} (1980) 431; Yad.\ Fiz.\ {\bf 32} (1980)
  838].


\bibitem{CFN}
Y.~M.~Cho,
  Phys.\ Rev.\  D {\bf 21} (1980) 1080;\\
L.~D.~Faddeev and A.~J.~Niemi,
  Phys.\ Rev.\ Lett.\  {\bf 82} (1999) 1624
  [arXiv:hep-th/9807069];\\
K.-I.~Kondo,
  Phys.\ Rev.\  D {\bf 74} (2006) 125003
  [arXiv:hep-th/0609166];\\
  K.-I.~Kondo, T.~Shinohara and T.~Murakami,
  arXiv:0803.0176 [hep-th].

\bibitem{lahiri}
T.~J.~Allen, M.~J.~Bowick and A.~Lahiri,
  Mod.\ Phys.\ Lett.\  A {\bf 6} (1991) 559;\\
A.~Lahiri,
  arXiv:hep-th/9301060;\\
  Phys.\ Rev.\  D {\bf 55} (1997) 5045
  [arXiv:hep-ph/9609510];\\
  Phys.\ Rev.\  D {\bf 63} (2001) 105002
  [arXiv:hep-th/9911107];\\
D.~S.~Hwang and C.~Y.~Lee,
  J.\ Math.\ Phys.\  {\bf 38} (1997) 30
  [arXiv:hep-th/9512216].


\bibitem{MacMan}
S.~W.~MacDowell and F.~Mansouri,
  Phys.\ Rev.\ Lett.\  {\bf 38} (1977) 739
  [Erratum-ibid.\  {\bf 38} (1977) 1376].


\bibitem{BF}
J. F. Plebanski,
  J.\ Math.\ Phys.\, {\bf 12} (1977) 2511;\\
J.~C.~Baez,
  Lect.\ Notes Phys.\  {\bf 543} (2000) 25
  [arXiv:gr-qc/9905087];\\
T.~Thiemann,
  Lect.\ Notes Phys.\  {\bf 631} (2003) 41
  [arXiv:gr-qc/0210094].


\bibitem{DT}
S.~Deser and C.~Teitelboim,
  Phys.\ Rev.\ D {\bf 13} (1976) 1592.


\bibitem{WY}
T.~T.~Wu and C.~N.~Yang,
  Phys.\ Rev.\  D {\bf 12} (1975) 3845;\\
D.~Z.~Freedman and R.~R.~Khuri,
  Phys.\ Lett.\  B {\bf 329} (1994) 263
  [arXiv:hep-th/9403031].


\bibitem{H}
M.~B.~Halpern,
  Phys.\ Rev.\ D {\bf 16} (1977) 1798.


\bibitem{GS}
O.~Ganor and J.~Sonnenschein,
  Int.\ J.\ Mod.\ Phys.\ A {\bf 11} (1996) 5701
  [arXiv:hep-th/9507036].
 

\bibitem{Schaden:1989pz}
  M.~Schaden, H.~Reinhardt, P.~A.~Amundsen and M.~J.~Lavelle,
  Nucl.\ Phys.\  B {\bf 339} (1990) 595.


\bibitem{KG}
T.~Kunimasa and T.~Goto, 
  Prog.\ Theor.\ Phys.\ {\bf 37} (1967), 452.
  

\bibitem{Stueckel}
E.~C.~G.~St\"uckelberg, 
  Annals Phys.\, {\bf 21} (1934) 367-389 and 744;\\
R.~Banerjee and J.~Barcelos-Neto,
  Nucl.\ Phys.\  B {\bf 499} (1997) 453
  [arXiv:hep-th/9701080].


\bibitem{Kondo:1996qa}
  K.~I.~Kondo,
  Prog.\ Theor.\ Phys.\  {\bf 98} (1997) 211
  [arXiv:hep-th/9603151].


\bibitem{BFV}
I.~A.~Batalin and E.~S.~Fradkin,
  Phy.\ Lett.\ B {\bf 180} (1986) 157;\\
  Nucl.\ Phys.\ B {\bf 279} (1987) 514;\\
E.~S.~Fradkin and G.~A.~Vilkovisky,
  Phys.\ Lett.\ B {\bf 55} (1975) 224;\\
I.~A.~Batalin and E.~S.~Fradkin, 
  Riv.\ Nuovo Cimento {\bf 9} (1986) 1;\\
M.~Henneaux,
  Phys.\ Rep.\ {\bf 126} (1985) 1.;\\
M.~Henneaux and C.~Teitelboim, {\it Quantization of Gauge Systems},
(Princeton Univ.\ Press, 1992);\\
T.~Fujiwara, Y.~Igarashi, and J.~Kubo,
  Nucl.\ Phys.\ B {\bf 341} (1990) 695;\\
  Phys.\ Lett.\ B {\bf 261} (1990) 427;\\
K.-I.~Kondo,
  Nucl.\ Phys.\ B {\bf 450} (1995) 251.


\bibitem{FT}
Freedman and Townsend,
  Nucl.\ Phys.\ B {\bf 177} (1981) 282-296.


\bibitem{KR}
K.~Hayashi,
  Phys.\ Lett.\ B {\bf 44} (1973) 497;\\
M.~Kalb and P.~Ramond,
  Phys.\ Rev.\ D {\bf 9} (1974) 2273;\\
E.~Cremmer and J.~Scherk,
  Nucl.\ Phys.\ B {\bf 72} (1974) 117;\\
Y.~Nambu,
  Phys.\ Rept.\ {\bf 23} (1976) 250;\\
P.~K.~Townsend,
  Phys.\ Lett.\ B {\bf 88} (1979) 97.


\bibitem{SO}
K.~Seo, M.~Okawa, and A.~Sugamoto, 
  Phys.\ Rev.\ D {\bf 12} (1979) 3744;\\
K.~Seo and M.~Okawa, 
  Phys.\ Rev.\ D {\bf 6} (1980) 1614.


\bibitem{TC}
S.~Weinberg,
Phys.\ Rev.\ D {\bf 19}, 1277 (1979);
L.~Susskind,
Phys.\ Rev.\ D {\bf
20}, 2619 (1979).


\bibitem{Dietrich:2006cm}
  D.~D.~Dietrich and F.~Sannino,
  Phys.\ Rev.\  D {\bf 75} (2007) 085018
  [arXiv:hep-ph/0611341].


\bibitem{walk}
B.~Holdom, 
Phys.\ Rev.\ D {\bf 24}, 1441 (1981).
K.~Yamawaki, M.~Bando and K.~i.~Matumoto,
Phys.\ Rev.\ Lett.\ {\bf 56}, 1335 (1986).
T.~W.~Appelquist, D.~Karabali and L.~C.~R.~Wijewardhana,
Phys.\ Rev.\ Lett.\ {\bf 57}, 957 (1986).
  V.~A.~Miransky and K.~Yamawaki,
  Phys.\ Rev.\ D {\bf 55}, 5051 (1997)
  [Erratum-ibid.\ D {\bf 56}, 3768 (1997)]
  [arXiv:hep-th/9611142].
  V.~A.~Miransky, T.~Nonoyama and K.~Yamawaki,
  Mod.\ Phys.\ Lett.\ A {\bf 4}, 1409 (1989).
K.~D.~Lane and E.~Eichten,
Phys.\ Lett.\ B {\bf 222}, 274 (1989).
E.~Eichten and K.~D.~Lane,
Phys.\ Lett.\ B {\bf 90}, 125
(1980).


\bibitem{Dietrich:2005jn}
  D.~D.~Dietrich, F.~Sannino and K.~Tuominen,
  Phys.\ Rev.\  D {\bf 72} (2005) 055001
  [arXiv:hep-ph/0505059];
  Phys.\ Rev.\  D {\bf 73} (2006) 037701
  [arXiv:hep-ph/0510217].


\bibitem{Foadi:2007ue}
  R.~Foadi, M.~T.~Frandsen, T.~A.~Ryttov and F.~Sannino,
  Phys.\ Rev.\  D {\bf 76} (2007) 055005
  [arXiv:0706.1696 [hep-ph]].


\bibitem{Appelquist:1999dq}
  T.~Appelquist, P.~S.~Rodrigues da Silva and F.~Sannino,
  Phys.\ Rev.\  D {\bf 60} (1999) 116007
  [arXiv:hep-ph/9906555].
  Z.~y.~Duan, P.~S.~Rodrigues da Silva and F.~Sannino,
  Nucl.\ Phys.\  B {\bf 592} (2001) 371
  [arXiv:hep-ph/0001303].


\bibitem{Diakonov}
D.~Diakonov and V.~Petrov,
  Grav.\ Cosmol.\  {\bf 8} (2002) 33
  [arXiv:hep-th/0108097].


\bibitem{tHooft}
G.~'t Hooft,
  Phys.\ Rev.\ D {\bf 14} (1976) 3432
  [Erratum-ibid.\ D {\bf 18} (1978) 2199].


\bibitem{CA}
see, e.g.
G.~Alexanian, R.~MacKenzie, M.~B.~Paranjape and J.~Ruel,
  arXiv:hep-th/0609146.


\bibitem{Gribov:1977wm}
  V.~N.~Gribov,
  Nucl.\ Phys.\  B {\bf 139} (1978) 1.


\bibitem{lunev}
F.~A.~Lunev,
  Mod.\ Phys.\ Lett.\  A {\bf 9} (1994) 2281
  [arXiv:hep-th/9407175];\\
  J.\ Math.\ Phys.\  {\bf 37} (1996) 5351
  [arXiv:hep-th/9503133];\\
M.~Bauer, D.~Z.~Freedman and P.~E.~Haagensen,
  Nucl.\ Phys.\  B {\bf 428} (1994) 147
  [arXiv:hep-th/9405028].


\bibitem{Harikumar:1996sf}
  E.~Harikumar and M.~Sivakumar,
  Phys.\ Rev.\  D {\bf 57} (1998) 3794
  [arXiv:hep-th/9604181].


\bibitem{Frohlich:1980gj}
  J.~Frohlich, G.~Morchio and F.~Strocchi,
  Phys.\ Lett.\  B {\bf 97} (1980) 249.


\end{thebibliography}
\end{document}